\documentclass[reprint,aps,superscriptaddress,twocolumn,showpacs,nofootinbib,longbibliography]{revtex4-2}

\usepackage{graphicx,color}
\usepackage{amsmath,amsfonts,enumerate,amsthm,amssymb,bbm}
\usepackage{float}
\usepackage[colorlinks=true,citecolor=blue,linkcolor=magenta]{hyperref}
\usepackage{multirow}
\usepackage{bbold}
\usepackage{braket}
\usepackage{soul}
\usepackage{svg}
\setsvg{inkscape=inkscape -z -D,svgpath=fig/}

\usepackage[version=4]{mhchem} 
\usepackage[caption = false]{subfig}

\usepackage{scrextend}
\usepackage{xcolor}
\usepackage{wrapfig}
\usepackage{cleveref}

\usepackage[switch,columnwise]{lineno}

\long\def\ca#1\cb{}

\newcommand{\abs}[2][]{#1| #2 #1|}

\renewcommand{\leq}{\leqslant}

\newcommand{\thv}{\vec{\theta}}

\newcommand{\alv}{\vec{\alpha}}
\newcommand{\gamv}{\vec{\gamma}}

\renewcommand{\vec}[1]{\boldsymbol{#1}}

\newcommand{\beginsupplement}{%
        \setcounter{table}{0}
        \renewcommand{\thetable}{S\arabic{table}}%
        \setcounter{figure}{0}
        \renewcommand{\thefigure}{S\arabic{figure}}%
     }
{}
{}

\begin{document}
\title{High-fidelity dimer excitations using quantum hardware}
\author{Norhan M. Eassa}
\affiliation{Department of Physics and Astronomy, Purdue University, West Lafayette, IN 47906, USA}
\affiliation{Quantum Science Center, Oak Ridge National Laboratory, Oak Ridge, TN 37831, USA}
\author{Joe Gibbs}
\affiliation{Department of Physics, University of Surrey, Guildford, GU2 7XH, UK}
\affiliation{AWE, Aldermaston, Reading, RG7 4PR, UK}
\author{Zoe Holmes}
\email[]{zoe.holmes@epfl.ch}
\affiliation{Institute of Physics, Ecole Polytechnique Fédérale de Lausanne (EPFL), CH-1015 Lausanne, Switzerland}
\affiliation{Information Sciences, Los Alamos National Laboratory, Los Alamos, NM 87545, USA}\affiliation{Quantum Science Center, Oak Ridge National Laboratory, Oak Ridge, TN 37831, USA}
\author{Andrew Sornborger}
\affiliation{Information Sciences, Los Alamos National Laboratory, Los Alamos, NM 87545, USA}
\affiliation{Quantum Science Center, Oak Ridge National Laboratory, Oak Ridge, TN 37831, USA}
\author{Lukasz Cincio}
\affiliation{Theoretical Division, Los Alamos National Laboratory, Los Alamos, NM 87545, USA}
\affiliation{Quantum Science Center, Oak Ridge National Laboratory, Oak Ridge, TN 37831, USA}
\author{Gavin Hester}
\affiliation{Department of Physics and Astronomy, Purdue University, West Lafayette, IN 47906, USA}
\affiliation{Department of Physics, Brock University, St. Catharines, ON, L2S3A1, Canada}
\author{Paul Kairys}
\affiliation{Mathematics and Computer Science Division, Argonne National Laboratory, Lemont, IL, 60439, USA} 
\author{Mario Motta}
\affiliation{IBM Quantum, IBM Research – Almaden, San Jose, CA 95120, USA}
\affiliation{Quantum Science Center, Oak Ridge National Laboratory, Oak Ridge, TN 37831, USA}
\author{Jeffrey Cohn}
\email[]{jeffrey.cohn@ibm.com}
\affiliation{IBM Quantum, IBM Research – Almaden, San Jose, CA 95120, USA}
\affiliation{Quantum Science Center, Oak Ridge National Laboratory, Oak Ridge, TN 37831, USA}
\author{Arnab Banerjee}
\email[]{arnabb@purdue.edu}
\affiliation{Department of Physics and Astronomy, Purdue University, West Lafayette, IN 47906, USA}
\affiliation{Quantum Science Center, Oak Ridge National Laboratory, Oak Ridge, TN 37831, USA}

\begin{abstract}

Many-body entangled quantum spin systems exhibit emergent phenomena such as topological quantum spin liquids with distinct excitation spectra accessed in inelastic neutron scattering (INS) experiments. Here we simulate the dynamics of a quantum spin dimer, the basic quantum unit of emergent many-body spin systems. While canonical Trotterization methods require deep circuits precluding long time-scale simulations, we demonstrate 'direct' Resource-Efficient Fast-forwarding (REFF) measurements with short-depth circuits that can be used to capture longer time dynamics on quantum hardware. The temporal evolution of the 2-spin correlation coefficients enabled the calculation of the dynamical structure factor $S(\mathbf{Q},\omega)$ - the key component of the neutron scattering cross-section. We simulate the triplet gap and the triplet splitting of the quantum dimer with sufficient fidelity to compare to experimental neutron data. Our results on current circuit hardware pave an important avenue to benchmark, or even predict, the outputs of the costly INS experiments.

\end{abstract}

\maketitle
\section{Introduction}
 Quantum spin systems show great potential to demonstrate exciting phenomena such as the emergent quantum anomalous Hall (QAH) effect and topological magnetoelectric effect \cite{lupke2022local}. Moreover, they serve as important candidates for topological materials that can potentially be used in dissipationless topological electronics and topologically protected computation \cite{hasan2010colloquium, he2019topological, bi2022drastic, serrano2022magnetic,chiesa2019quantum,moreno2021measuring}. Of particular importance are quantum spin liquids (QSL) which are highly entangled many-body spin systems. Originally proposed by P.W. Anderson in 1973 \cite{anderson1973resonating}, these systems are breeding grounds for emergent quasi-particles exhibiting spin fractionalization and highly non-trivial topological characteristics \cite{balents2010spin}. In the quantum computing community, QSLs have created a recent interest as excitations of spin liquid states could be used to test the efficacy of quantum hardware to produce truly long-range entangled states \cite{ebadi2021quantum,chiesa2019quantum}, as well as novel ways to create error-resilient ground and excited states \cite{richter2021simulating, kalinowski2022non,banerjee2017neutron}. 
 
 The smallest unit of such an entangled spin state is the quantum dimer - a single valence-bond unit between two $s = 1/2$ spins. The canonical proposal for a QSL required a coherent interaction between several such dimer spin pairs called 'dimer coverings' \cite{anderson1993resonating}. As a result, the ground state of these QSLs consists of entangled quasi-particle states where the spins lose their individual character and are instead defined by the $s=0$ singlet state $\ket{\uparrow\downarrow-\downarrow\uparrow}/\sqrt{2}$ ($s_z = 0$) and $s=1$ triplet states ${\ket{\uparrow\uparrow}}$ ($s_z = 1$), $\ket{\uparrow\downarrow+\downarrow\uparrow}/\sqrt{2}$ $(s_z = 0)$, and ${\ket{\downarrow\downarrow}}$ ($s_z = -1$) \cite{urushihara2020crystal} separated by a gap $\Delta$ - the spin-triplet gap. 
 
 \begin{figure}[htp!]
    
    \centering
    
    \includegraphics[width=\columnwidth]{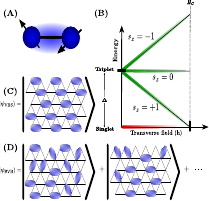}
    
    \caption{\textbf{The physics of dimers}: \textbf{(A)} Cartoon of a spin dimer, consisting of a single valence-bond unit between two $s = 1/2$ spins. \textbf{(B)} Energy gap between singlet and triplet states of the above dimer in the presence of a transverse field $h$, represented by $\Delta$. The transverse field also splits the once degenerate triplet state into 3 different energy levels up until the critical magnetic field $H_C$. \textbf{(C)} Cartoon of a VBS state on a triangular lattice, shown as a specific pattern of long-range ordered entangled pairs of spins. These entangled pairs are indicated by the dimer coverings (blue ovals) that cover two sites on the triangular lattice. \textbf{(D)} In an RVB state, the wavefunction is a superposition of numerous different pairings of spins. The valence bonds can be short-range or long-range.}
    \label{VBS}
\end{figure} 

 With the application of a magnetic field, the triplet states split linearly proportional to the field into three separate levels leading to three distinct spin gaps relative to the ground state. 
 
There is a continuing interest in ground states that host such spin valence bonds, including Valence Bond Solids (VBS) (long-range ordered state of dimers) and Resonant Valence Bond (RVB) liquids (long-range entangled dimers), stabilized in a soup of frustrated lattices. A detailed discussion of the recent research in these systems is available in Broholm et al., \cite{broholm2020quantum}, including their possible future contributions to enhancing quantum computing hardware. 

While several material candidates are proposed to host such states \cite{chaloupka2010kitaev,veiga2017pressure, plumb2014alpha, takagi2019concept, banerjee2016proximate,balents2010spin, haravifard2014emergence}, their ground states being devoid of any net moment makes their identification experimentally difficult. The signatures of quantum magnets are often revealed by their excitations. Inelastic Neutron Scattering (INS) is a very common avenue to probe these dynamics for several reasons \cite{furrer2009neutron, broholm2020quantum}. The neutrally-charged neutrons do not interact with the electrons directly. Many isotopes have a large neutron penetration depth making them suitable for probing bulk properties of matter. Fast neutrons from spallation and fission can be slowed using moderators to energies comparable to the elementary excitation energy in magnetic matter \cite{adams2020first}. The resulting de Broglie wavelength is on the order of the lattice constant of many condensed matter single crystals. The neutron's magnetic moment can scatter from the spins and directly access the dynamic 2-spin correlations of quantum fluctuating spin systems. All of the above makes INS a staple probe for the study of the energy levels of spin excitations in magnetic matter. 

The simulation of quantum spin Hamiltonians on a quantum computer is a rapidly improving avenue for the simulation of larger, more complicated, quantum spin systems without the fermionic sign problem \cite{ortiz2001quantum, bravyi2002fermionic, huggins2022unbiasing}. Such simulations open new avenues to benchmark both the fidelity of the energy levels and the clarity of the spin gaps. Longer term, such quantum simulations may allow one to predict the output of INS experiments, providing a means of minimizing the number of costly INS experiments that need to be performed. Another exciting possibility is for the simulation of the dimer, and inter-dimer interactions of 2-spin qubit systems such as those realized in Diamond lattices studied with NV-Centers \cite{bartling2022entanglement}. However, noise levels on current quantum computers make it hard to simulate the quantum spin dimer in a magnetic field. Here we investigate methods to sidestep the barriers posed by noise on current hardware.

 The zero-temperature magnetic neutron cross-section (i.e., intensity of scattered neutrons from the magnetic moments in a solid) \cite{lovesey1984theory} produces an intensity spectrum that can be written as 
\begin{equation}
\begin{split}\label{eq:IntensitySpectrum}
    I \equiv & \frac{d^2 \sigma}{d \Omega d E_p} =  r_0 ^2 \frac{k_f}{k_i} \left|\frac{g}{2} F(Q)\right|^2 \sum_{\alpha , \beta = x,y,z} (\delta_{\alpha , \beta} - \hat{Q}_\alpha \hat{Q}_\beta)\\
    & \times \frac{1}{2\pi \hbar} \int^{\infty}_{-\infty} dt\hspace{3pt} e^{-i \omega t } \frac{1}{N}\sum_{i,j} \langle s^{\alpha} _i (t) s^{\beta} _j (0) \rangle_0  e^{-i\mathbf{Q}\cdot \mathbf{R}_{ij}} \, , 
\end{split}
\end{equation}
where $\mathbf{Q}$ is a momentum vector, $k_i$ and $k_f$ are the wavevectors of incident and scattered neutrons respectively, $t$ is time, $\omega$ is frequency, $r_0$ is the inter-atomic distance, $g$ is the gyromagnetic ratio, $F(Q)$ is the magnetic form-factor which depends on the isotope, $N$ is the number of unit cells in the solid, and $R_{ij}$ are the relative positions of the magnetic ions with spin components $s^\alpha _i$ \cite{furrer2009neutron}. 

The magnetic scattering cross section for the INS intensity is proportional to a quantity called the dynamic structure factor $S^{\alpha,\beta}(\mathbf{Q},\omega)$, given by
\begin{equation}\label{DFS}
    \begin{split}
        S^{\alpha,\beta}(\mathbf{Q},\omega) =  \frac{1}{2\pi\hbar} \int^{\infty}_{-\infty} dt\hspace{3pt} e^{-i \omega t } &\\
        \times \frac{1}{N} \sum_{i,j} \langle s^{\alpha} _i (t) s^{\beta} _j (0) \rangle_0  e^{-i\mathbf{Q}\cdot \mathbf{R}_{ij}} .
    \end{split}
\end{equation}
Hence, $I \propto \sum_{\alpha,\beta} (\delta_{\alpha,\beta} - \hat{Q}_\alpha \hat{Q}_\beta)S^{\alpha,\beta}(\mathbf{Q},\omega)$.

At the core of this expression are the coefficients of the dynamical spin-spin correlation functions,
\begin{equation}
\begin{split}\label{eq:CorrelationFunc}
    C^{\alpha \beta} _{i,j} (t)  &:=  \langle  s^{\alpha} _i (t) s^{\beta} _j (0)  \rangle_{0} \\&=  \sum_{p} \langle{0|s^\alpha_{i}|p}\rangle \langle{p|s^\beta_{j}}|0\rangle e^{-iE_{p} t} \, ,
\end{split}
\end{equation}
where $|0\rangle$ and $|p\rangle$ are the ground and excited molecular eigenstates with energies $E_0$ and $E_p$ respectively. The dynamic structure factor is interpreted to be the space-time Fourier transform of the correlation functions. The factor of $e^{-i E_p t}$ makes the connection to the specific system Hamiltonian used to construct the time evolution operator $e^{-i H t}$, whose Fourier transform gives the power spectrum. For a real INS experiment, depending on the experimental conditions such as the polarization of the neutron beam and the direction of the $\mathbf{Q}$ momentum vector with respect to the anisotropy direction of the dimer, these coefficients are weighted by  $(\delta_{\alpha , \beta} - \hat{Q}_\alpha \hat{Q}_\beta)$ which represents a selection rule that only the scattering components perpendicular to the $\mathbf{Q}$ momentum vector contribute to the intensity.  

\indent Thus, our central task is to compute the coefficients of the dynamical spin-spin correlation functions
\begin{equation}
\begin{split}\label{eq:CorrelationFuncExpl}
    C^{\alpha \beta} _{i,j} (t)  &:=  \langle 0 | e^{+i H t} \sigma^{\alpha} _i  e^{-i H t} \sigma^{\beta}_j | 0 \rangle  \, .
\end{split}
\end{equation} The canonical approach to calculate the time evolution on a gate-based quantum computer is Trotterization \cite{suzuki1993improved,sornborger1999higher}. 
On a fault-tolerant quantum computer, Eq.\eqref{eq:CorrelationFuncExpl} may be computed using the Hadamard test (the so-called \textit{indirect measurement} scheme), with time evolution $e^{-i H t}$ implemented using iterated Trotter steps \cite{suzuki1993improved,sornborger1999higher,chiesa2019quantum}, as illustrated in Fig.\ref{fig:VFFvsTrott}. However, this scheme leads to linear growth in circuit depth over time. Thus, current quantum hardware is incapable
of producing reliable dynamics for more than a few timesteps because the circuit depth extends beyond the coherence time of the quantum hardware.

\indent Here we explore state-of-the-art approaches to circumvent the effect of hardware noise. In particular, we explore: 
\begin{enumerate}[A)]
    \item  The 
    \textit{Resource Efficient Fast-Forwarding (REFF)} algorithm \cite{gibbs2022dynamical} to enable long-time simulations using a fixed depth circuit. 
    \item  
    The \textit{direct measurement scheme} which uses mid-circuit measurements instead of controlled operations between an ancillary qubit and system qubits. 
    \item \textit{Measurement error mitigation} to reduce the effect of noise on our final measurements.
\end{enumerate}

The details of these different methods are elaborated on in length in Section~\ref{meth}.

Our Hamiltonian of choice to study in this paper is the \textit{1-D Heisenberg model with a transverse field (XYZ+h)} for 2-spin 1/2 systems (i.e., dimers), which is written as follows:
    \begin{eqnarray}\label{heis_eq}
    H_{\rm Heis} & = &  \sum\limits_{j=1}^{n - 1} J_{xx}\sigma^{x}_{j}\sigma^{x}_{j+1} + J_{yy}\sigma^{y}_{j}\sigma^{y}_{j+1} + J_{zz}\sigma^{z}_{j}\sigma^{z}_{j+1} \nonumber \\
    && +
    h \sum\limits_{j=1}^{n} \sigma^{z}_{j} \, .
   \end{eqnarray}
where $\sigma^{x}_{j}$, $\sigma^{y}_{j}$ and $\sigma^{z}_{j}$ represent Pauli operators on the $j_{\rm th}$ qubit.

\indent 
In this manuscript, we concentrate on near-term quantum algorithms for computing the dynamic structure factor from inelastic neutron scattering for the 2-spin correlation function for a magnetic dimer. Using REFF we calculate the coefficients of the dynamical spin-spin correlation functions of 2 spin-1/2 systems under the action of different Hamiltonian models. Computing spectral estimates of the temporal correlation coefficients, we obtain INS intensity spectra under the action of each Hamiltonian model. We also show significant improvement in solution quality using direct (ancilla-free) measurements. Our results, validated on IBM-Q backends, represent a multi-fold improvement in the fidelity of the simulated results of a 2-spin dimer on quantum hardware. Besides the spin-triplet gap, we were able to clearly observe the splitting of triplet states under the action of a transverse field. We argue that the fidelity of our techniques is sufficient for the comparison of the spectrum to experimental data.
\section{Methods}\label{meth}

\subsection{Inelastic neutron scattering calculations}\label{insCalc}

To determine the general two-point two-site correlation functions for a dimer, we have to measure the correlation functions $C^{\alpha\beta}_{i,j}(t)$ for all $\alpha,\beta = x,y,z$ and for all $i,j = 1,2$. Thus, there are 36 coefficients in total that need to be measured on quantum hardware and combined within the calculation of Eq.\eqref{DFS}.

\subsection{Simulation methods}

\begin{figure}[htp!]
    \centering
    \includegraphics[width =\columnwidth]{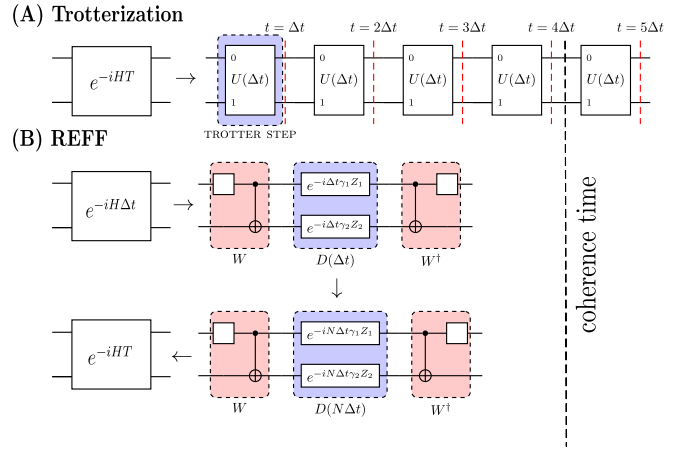}
    \caption{\textbf{The concept of REFF}: \textbf{(A)} An example of a Trotterization-based quantum simulation with $N = 5$ time steps. Such a simulation runs past the coherence time of the qubits of the physical architecture for a sufficiently precise result. \textbf{(B)} An example of a REFF-based quantum simulation. An approximate diagonalization of a short-time Trotterization unitary is found variationally in the form of the ansatz $V_{\Delta t}(\boldsymbol{\alpha}) = W(\boldsymbol{\theta}) D_{\Delta t}(\boldsymbol{\gamma}) W^\dagger (\boldsymbol{\theta})$, as explained in Fig.\ref{flowchart}. Such ansatz allows for constant circuit depth when simulating the time evolution as only the parameters of $D_{\Delta t}(\boldsymbol{\gamma})$ need to be modified accordingly. Hence, if the sequence of gates forming our ansatz is small enough to not exceed the coherence time, longer time simulations can be performed than in the case of Trotterization.}
    \label{fig:VFFvsTrott}
\end{figure}

\paragraph*{Trotterization:}
In our implementations of the iterated Trotter method, we use a first-order Trotterization. That is, if we were to have a local Hamiltonian
\begin{equation}\label{eq:trotterstep0}
 H = \sum^{k}_{j = 1} H_j \,  ,
\end{equation} 
where the $k$ component Hamiltonians $H_j$ represent the individual terms in the Hamiltonian $H$, we approximate the short-time evolution operator of the system with \cite{knee2015optimal}:
\begin{equation}\label{eq:trotterstep}
  U(\Delta t) = e^{-iH \Delta t} \approx \prod^{k}_{j = 1} e^{-iH_j \Delta t} \,  ,
\end{equation} 
 Since the Hamiltonians we consider are local, each of the terms $e^{-iH_j \Delta t}$ act non-trivially on at most two qubits, and thus it is straightforward to compile a circuit to implement the right hand side of Eq.\eqref{eq:trotterstep}. The longer time evolution of the system can then be approximated by applying this circuit iteratively. That is, we have
\begin{equation}\label{eq:trotterstep-iter}
   U(t) = e^{-iH N \Delta t} \approx \left( \prod^{k}_{j = 1} e^{-iH_j \Delta t} \right)^N \, ,
\end{equation}
and so to simulate up to time $t = N \Delta t$, we simply apply the initial gate sequence for approximating the short-time evolution (known as the trotter step) $N$ times as sketched in Fig.\ref{fig:VFFvsTrott}.

\paragraph*{Resource Efficient Fast-forwarding (REFF):}
On current quantum hardware, the length of time that can be viably simulated via Trotterization-based simulation is limited by the build-up of gate errors and the short coherence times of the available devices. Variational fast-forwarding algorithms \cite{cirstoiu2020variational, commeau2020variational,gibbs2022long,geller2021experimental,gibbs2022dynamical}, have been introduced as a means to sidestep this limitation of near-term hardware. This family of algorithms employs a hybrid quantum-classical optimization loop to learn a fixed-depth circuit that can be used to simulate arbitrary times. Hence, as sketched in Fig.\ref{fig:VFFvsTrott}, these methods have the potential to increase the lengths of times that may be simulated within the finite coherence times of current devices. 

Here we focus on the Resource Efficient Fast-forwarding (REFF) algorithm \cite{gibbs2022dynamical}. This algorithm variationally searches for an approximate diagonalization of the Trotter approximation of the short-time evolution of the system, i.e., Eq.\eqref{eq:trotterstep}. That is, we use an ansatz of the form 
\begin{align}\label{eq:VHDansatz}
V_{t}(\alv)= W(\thv)D_t(\gamv)W^\dagger(\thv) \,,
\end{align}
where $\alv = (\thv, \gamv)$, $D_t(\gamv)$ is a time-dependent unitary that is diagonal in the standard basis and $W(\thv)$ is a time-independent unitary encoding a rotation into the eigenbasis of $H$. Since $D_t$ is diagonal, $N$ applications of $D_{t}$ are equivalent to one application of $D_{Nt}$, i.e., $D_{t}(\gamv)^N = D_{N t}(\gamv)$. Hence if we find a set of optimized parameters $\alv_{\rm opt} = (\thv_{\rm opt} , \gamv_{\rm opt})$ such that $V_{t}(\alv) \approx U(\Delta t)$ then we have that
\begin{align}\label{eq:VHDsim}
U(t)= W(\thv)D_{Nt}(\gamv)W^\dagger(\thv) \ .
\end{align}
Thus we can simulate arbitrary length times using a fixed-depth circuit.
The algorithm is inspired by quantum machine learning in the sense that it draws on recent analytic bounds from quantum learning theory \cite{caro2022generalization, caro2022outofdistribution} which shows that one can learn the diagonalization by studying its effect on only a polynomial number of unentangled random product states. The training is performed on a number of training states using the following cost function:
\begin{equation}
      C_{\text{REFF}}(\alv) := 1 - \frac{1}{N}  \sum^{N}_{j = 1} \big|\langle \psi_P^{(j)}| V(\alv)^\dagger U | \psi_P^{(j)} \rangle \big|^2 \, ,
\end{equation} 
where the training states are drawn from the distribution of products of single qubit Haar random states $| \psi_P^{(j)} \rangle \sim \text{Haar}_1^{\otimes n}$ and $N \in O(\text{poly}(n))$. The objective of such a training procedure is to find the optimal parameters $\boldsymbol{\alpha}_\text{opt} = (\boldsymbol{\theta}_\text{opt}, \boldsymbol{\gamma}_\text{opt})$ which are used to implement the fast-forwarded simulation $W(\thv)D_{Nt}(\gamv)W^\dagger(\thv)$, with the fast-forwarding error growing quadratically in the simulation time. A flowchart detailing the algorithm is shown in Fig.\ref{flowchart}.

\subsection{Measurement methods}
\begin{figure}[htp!]
    \centering
    \includegraphics[width=\columnwidth]{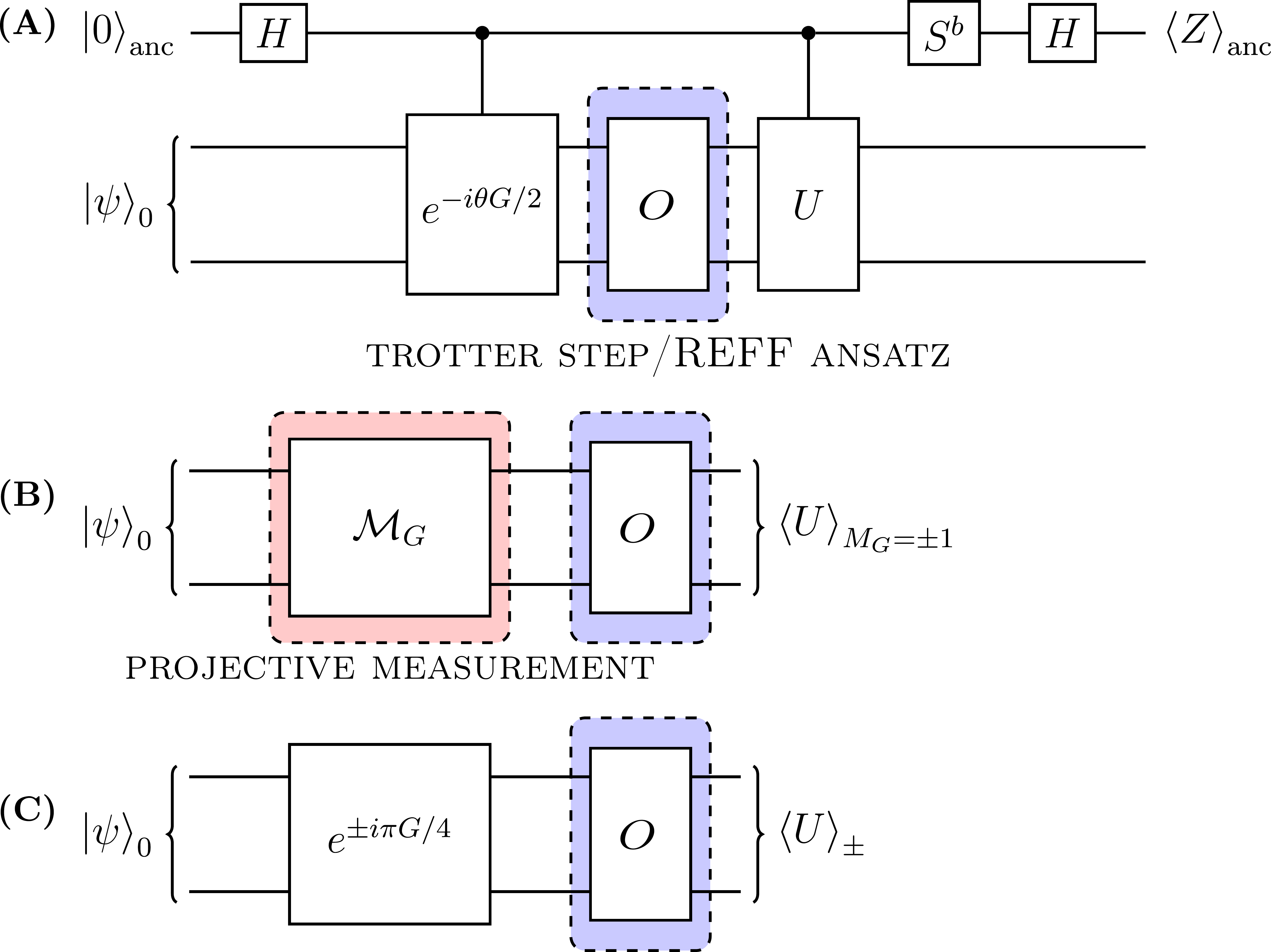}
    \caption{\textbf{Methodology to convert indirect measurements to direct measurements:} \textbf{(A)} Hadamard test with two controlled gates. $O$ can be any arbitrary quantum gate. In the case of the circuits used to produce the results in this paper, $O$ is the sequence of gates representing the time evolution operator $e^{-iHt}$ (i.e., the Trotterization unitary or REFF ansatz). $H$ is the Hadamard gate. $S$ is $e^{-i\pi Z/4}$ and $b\in \{0,1\}$. When $b = 0$, $\langle Z \rangle = \text{Re}(\langle \psi_{\text{in}}| W |\psi_{\text{in}}\rangle)$, and when $b = 1$, $\langle Z \rangle = \text{Im}(\langle \psi_{\text{in}}| W |\psi_{\text{in}}\rangle)$. In our case of calculating the coefficients of the dynamical spin-spin correlation functions, the gates $e^{i\theta G/2}$ and $U$ can be the $X, Y$, or $Z$ operator, depending on which coefficient is being calculated. With our choice of spin operators in place, along with the sequence of gates representing the time evolution operator, we are capable of measuring the real and imaginary parts of the coefficients, depending on the value of $b$, whether it be 0 or 1. \textbf{(B)},\textbf{(C)} Quantum circuits to estimate the output of \textbf{(A)} using direct measurements (i.e., without the use of an ancillary qubit and controlled operations). $\mathcal{M}_G$ is the projective measurement of $G$. In Sec.~\ref{DMeas}, it is illustrated how to get the corresponding measurements to the two cases of $b$ in \textbf{(A)}.}
    \label{fig:DirectIndirectMeth}
\end{figure}

When collecting the output from a quantum computation in the form of a measurement, there are two ways to go about it: indirect and direct measurements.
\paragraph*{Indirect measurements:}
The standard method of computing the coefficients of the form in Eq.\eqref{eq:CorrelationFuncExpl} (along with other measurements) is via a Hadamard test. This circuit measures $\langle \psi|O|\psi\rangle$ for any unitary $O$ by i. preparing an ancillary qubit in the $|+\rangle = \frac{1}{\sqrt{2}} ( |0 \rangle + | 1 \rangle )$ by applying a Hadamard gate, ii. applying a C-$U$ controlled gate to the state $\ket{\psi}$ conditional on the ancilla qubit being in the state $\ket{1}$ and iii. measuring the expectation value of the Pauli $Z$ operator on the ancillary qubit \cite{somma2002simulating}. This method is said to employ an 'indirect measurement' as the system register is not directly measured. Depending on our choice of $b$ as shown in Fig.\ref{fig:DirectIndirectMeth}, we measure the real or imaginary part of the coefficient.

\paragraph*{Direct measurements:}
As described in Ref.~\cite{mitarai2019methodology}, assuming $G$ is a Hermitian operator where $G^2 = I$, and $U$ is an arbitrary quantum gate that satisfies either of the following conditions:
\begin{itemize}
    \item $U$ can be decomposed into the sum of Pauli products with a polynomial number of terms with respect to the number of qubits.
    \item Let $k$ be an integer such that $k = \text{polylog}(n)$, where $n$ is the number of qubits. For any $k$-local quantum gate $U$, it is possible to estimate $\langle \psi| U | \psi \rangle$ up to the precision $\epsilon$ in time $\mathcal{O}(\frac{k^2 2^k}{\epsilon ^ 2})$ without the use of the Hadamard test.
\end{itemize}
the controlled operations can be converted into projected partial measurements between the system qubits, employing a ``direct measurement" on the system qubits. An illustration of this methodology is shown in Fig.\ref{fig:DirectIndirectMeth}. Hence, we can forgo the use of an ancilla and measure the system state directly. This reduces the number of 2-qubit gates required to compute the correlation function; however, comes at the cost of requiring mid-circuit measurements (an alternative source of noise). We further investigate which method proves more accurate on current hardware. 
\subsection{Measurement error mitigation}
Finally, we will employ measurement error mitigation to improve the quality of our final simulation. Noisy quantum computers can have a systematic bias when measuring a quantum state, causing the estimated probability amplitudes of the wavefunction to be skewed from the true values. Measurement error mitigation seeks to quantify and undo this bias to transform the measured counts to a better approximation of the ideal (noiseless) case.

Denoting the vector of probabilities that
would have been obtained from an ideal quantum computer as $p_{\rm ideal}$ and the actual probabilities obtained on the real noisy device as $p_{\rm noisy}$, we can write 
\begin{equation}
    p_\text{noisy} = M p_\text{ideal} \, ,
\end{equation}
where $M$ is a stochastic, invertible map capturing the effect of measurement noise \cite{maciejewski2020mitigation, Qiskit}. The matrix element $M_{ij}$ is calculated by measuring the probability that the computational basis state $|i\rangle$ is measured when the state $|j\rangle$ is prepared. When state preparation and measurement errors are absent, this matrix resolves to the identity. This map $M$ is known as a calibration matrix and can be (approximately) learned using measurement spectroscopy. By then applying the inverse of the calibration matrix to noisy probability vector, 
\begin{equation}
    p_\text{ideal} = M^{-1} p_\text{noisy} \, ,
\end{equation}
it is possible to construct a closer approximation of the ideal probabilities. 
\section{Results}

\begin{figure*}[htp!]
    \centering
    \includegraphics[width=1\textwidth]{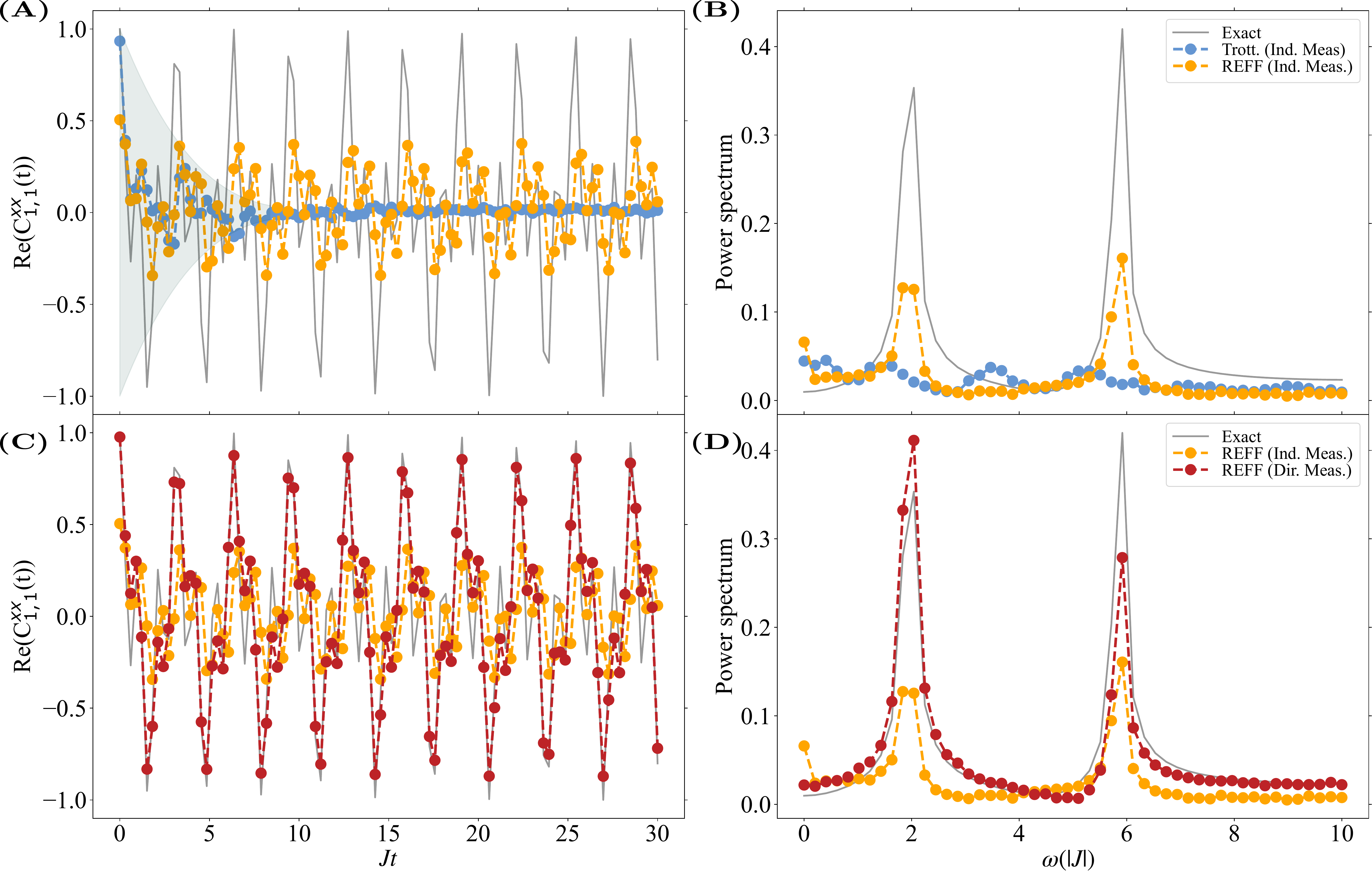}
    \caption{\textbf{Comparison of accuracy of different methods on the $C_{xx}$ correlation function:} In the different panels, results of measuring the real  part of the correlation function $C^{xx}_{1,1}(t)$ over time and its Fourier transform in frequency $\omega$ are presented respectively. The Hamiltonian employed is the Heisenberg model for a 2 spin-1/2 system.  \textbf{(A)},\textbf{(B)} Show both the Trotterization (light blue dots) and the REFF (light orange dots) results. The shaded area highlights the exponential decay of the correlation function over time when using Trotterization. On the other hand, we see how the trend of the REFF results stays mostly constant over time. In \textbf{(C)}, \textbf{(D)}, the difference between the implementation of indirect vs. direct measurements (dark red dots) is presented. We observe a significant increase in the accuracy of the results (especially in the amplitudes) when implementing direct measurements. Results were obtained using \textit{ibm\_geneva}, with the qubits' initial layout of [1,2] for the direct measurements and [0,1,2] for the indirect measurements, along with setting $t = 30$, the number of time steps $= 100$, $dt = 0.3$, and the number of shots $= 8000$.}
    \label{mit_ex}
\end{figure*}

\begin{figure*}[htp!]
    \centering
    \includegraphics[width=\textwidth]{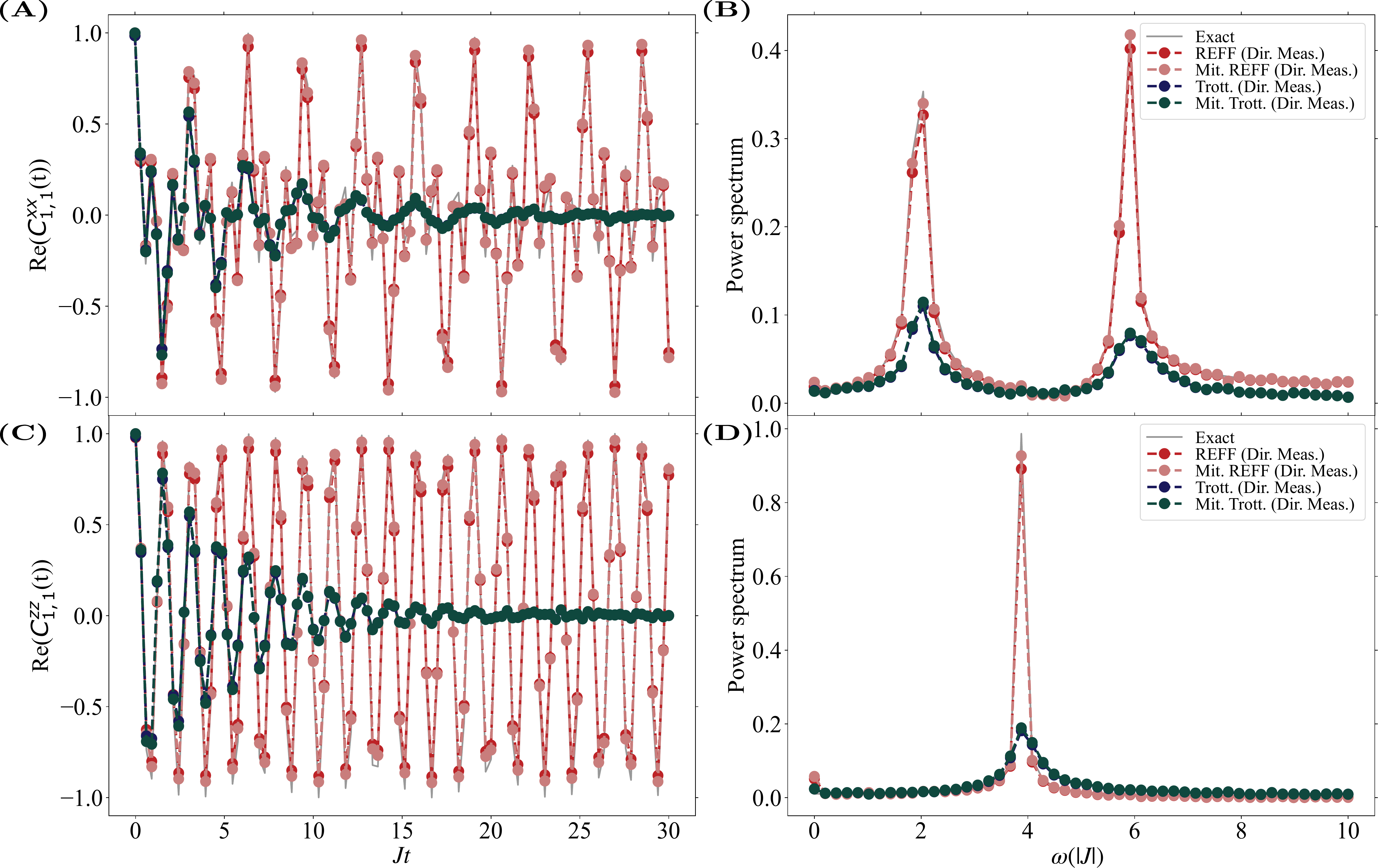}
    \caption{\textbf{Correlation functions and effect of using different hardware/qubits}: We set $J = 1$ and $h = 1$ for these results. Both Trotterization (dark blue dots, dark green dots for measurement error mitigated results) and REFF (dark red dots, pink dots for measurement error mitigated results) were used to simulate the time evolution, and direct measurements were implemented. The use of measurement error mitigation is also presented, showing modest gains in accuracy. \textbf{(A)} is the plot of the real part of the correlation functions $C^{xx}_{1,1}(t)$ against $(J = 1)t$, while \textbf{(B)} is the plot of its corresponding Fast Fourier transform (representing the power spectra) against frequency $\omega$. In \textbf{(C)} and \textbf{(D)}, we present the same results for the real part of the $zz$ correlation function, Re$(C^{zz}_{1,1}(t))$. The REFF results were of the measurements used in the total dynamical scattering factor calculation of the results in Fig.\ref{Heis_Int_Dir}. The results were obtained using \textit{ibm\_auckland}, with the qubits' initial layout of [4,1] for the direct measurements, along with setting $t = 30$, the number of time steps $= 100$, $dt = 0.3$, and the number of shots $= 8000$. Compared to the results in Fig.\ref{mit_ex}, these results demonstrate an increase in accuracy when using different hardware/qubits of better calibrations. Hence, we observe how different calibrations affect the output of our results. It is to be noted that the measurement of the real part (and imaginary part) of the correlation function $C^{yy}_{1,1}(t)$ is practically identical to that of $C^{xx}_{1,1}(t)$.}
    \label{fig5}
\end{figure*}

\begin{figure}[htp!]
\centering
       \includegraphics[width=1.0\columnwidth]{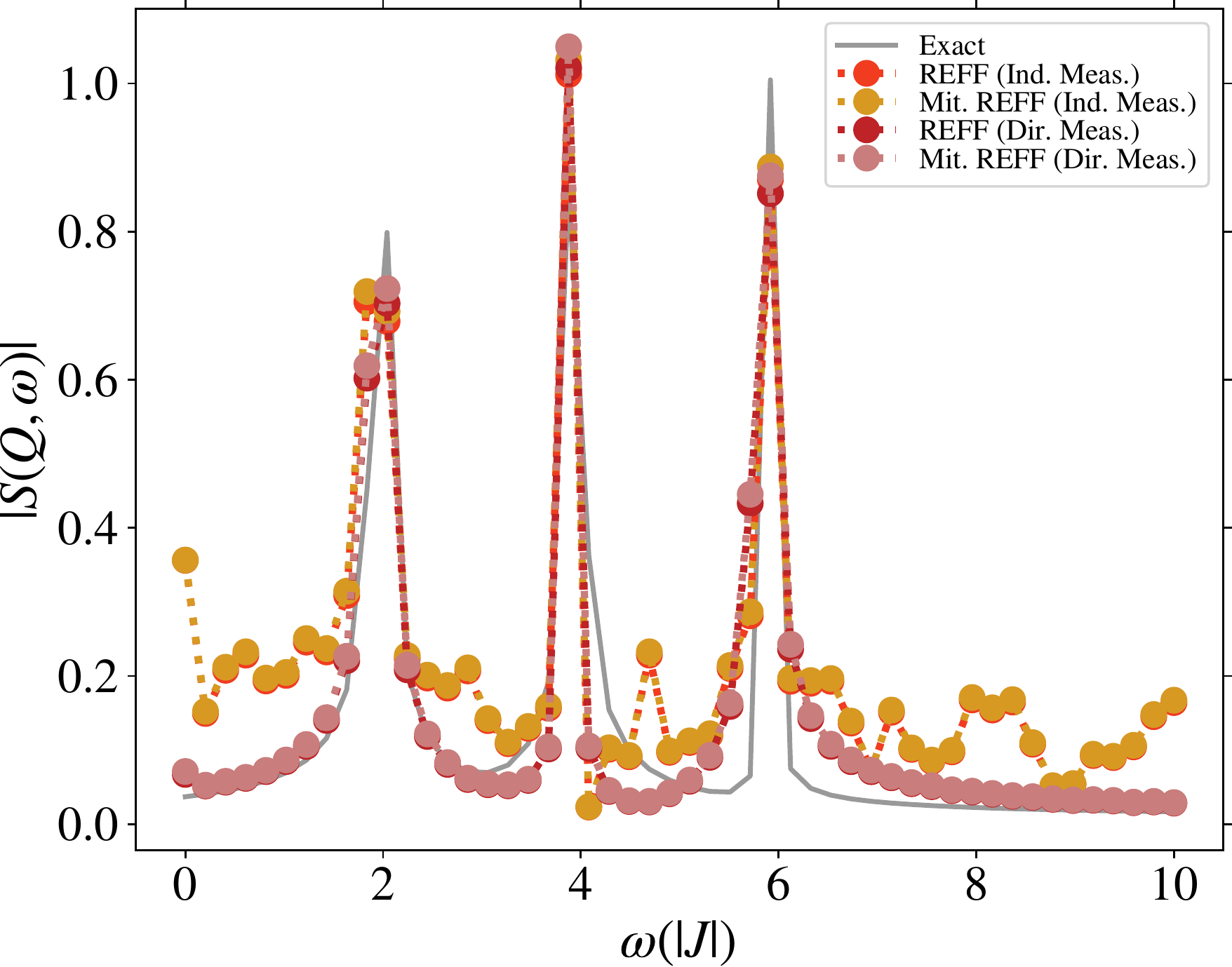}
\caption{\textbf{The spin-triplet splitting:} The dynamical scattering cross-section is calculated with the correlation functions measured on the IBM hardware as according to Eq.\eqref{DFS} and plotted against frequency $\omega$. REFF was utilized to simulate the time evolution. The results are for a dimer system with Heisenberg interaction ($J = 1$) and transverse field ($h = 1$) at \textbf{Q} = 0. The results were obtained using \textit{ibm\_auckland}, with the qubits' initial layout of [4,1] for the direct measurements (dark red dots, pink dots for measurement error mitigated results) and [7,4,1] for the indirect measurements (orange dots, yellow dots for measurement error mitigated results). For the simulations, $t = 30$, the number of time steps $= 100$, $dt = 0.3$, and the number of shots $= 8000$.}
\label{Heis_Int_Dir}
\end{figure}

\begin{figure}[htp!]
    \centering
    \includegraphics[width=\columnwidth]{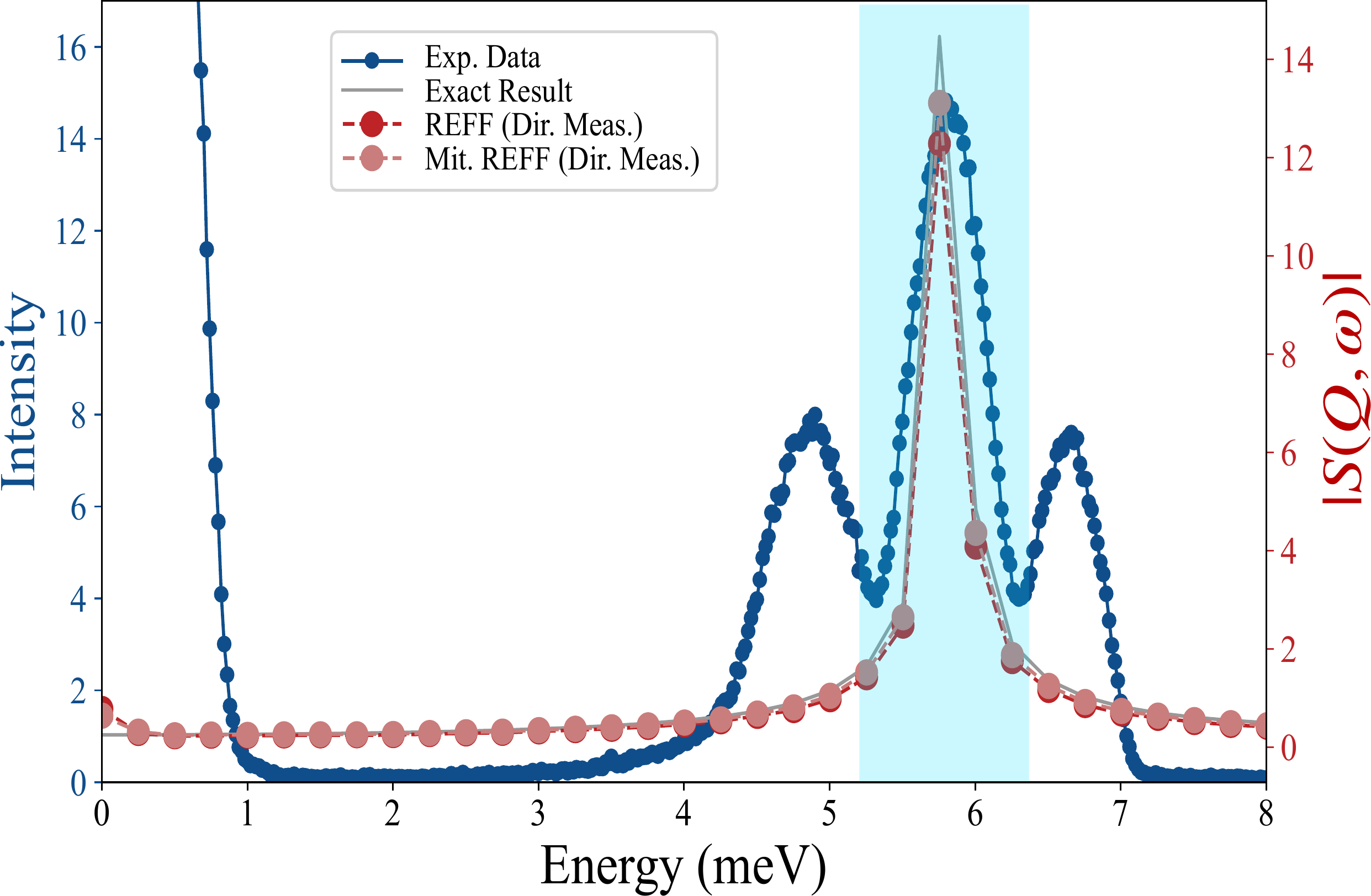}
    \caption{\textbf{A comparison to experiment}: We calculate and plot the dynamical scattering factor for the XY + ZZ perturbation Hamiltonian model presented in ref \cite{kurita2019localized} against energy in meV. The inelastic neutron scattering data is plotted (teal blue dots) against energy in meV with an energy intergation in $\mathbf{Q}$ from 0.9 to 1.1 $\AA^{-1}$. More details on this can be found in Ref. \cite{kurita2019localized}. We also plot the exactly calculated results (grey lines), direct REFF (dark red dots) measurements, and the measurement error mitigated (pink dots) measurements for the model proposed in \cite{kurita2019localized} which only accounts for the middle peak (blue shaded region). In the simulation, we obtained the middle peak (shaded area) - the two other peaks have been proposed to be from the unpaired single spin excitations \cite{kurita2019localized} due to the dimers' placement in the lattice, and hence unaccounted for in the model for the simulations. These results were obtained on \textit{ibm\_hanoi} on qubits [9,8] where only direct measurements were implemented. }
    \label{sim_Res}
\end{figure}

In this section, we demonstrate that the methods illustrated in this paper can be used to accurately compute the spin-spin correlation functions, required for inferring the magnetic scattering cross-section measured in INS experiments on quantum hardware. Figs.\ref{mit_ex} and \ref{fig5} include results for specific spin-spin correlation functions when setting $J = 1$ and $h = 1$ for the 2 spin-1/2 Heisenberg interaction system described in Eq.\eqref{heis_eq}. Fig.\ref{Heis_Int_Dir} then includes the result of calculating the complete dynamical structure factor from these spin-spin correlation functions as indicated from Eq.\eqref{DFS}. Lastly, Fig.\ref{sim_Res} includes a comparison between simulation results on IBM hardware to actual INS experimental data for an XY + ZZ perturbation dimer model as indicated from Eq.\eqref{eq:molmodel}.

\medskip

\paragraph*{Simulated correlation functions:}
In Fig.\ref{mit_ex}, we plot the real part of the $xx$ correlation function, Re$(C^{xx}_{1,1}(t))$, using both Trotterization and REFF, compared to the analytically computed exact results. Such results were obtained on \textit{ibm\_geneva}, with the qubits' initial layout set to [0,1,2] for indirect measurements and [1,2] for the direct measurements. The calibration details of the qubits are stated in Table \ref{table:2}. As shown in \textbf{(A)}, the correlation function computed via Trotterization (light blue dots) decays exponentially with simulation time and essentially vanishes by $Jt \sim 7$ (shaded envelope in Fig.\ref{mit_ex}\textbf{(A)}). This happens due to the linear growth in circuit depth with time for Trotterization. In \textbf{(B)}, we plot the Fourier transform of Re$(C^{xx}_{1,1}(t))$, which represents its power spectrum in the frequency domain.  The exact results show  that the correlation function $C^{xx}_{1,1}(t)$ has two peaks in frequency corresponding to $\omega = 2 |J|$  and $\omega = 6 |J|$. 
However, the indirect results with Trotterization on \textit{ibm\_geneva} fail to show any peak at the correct energies, instead showing undulations at ad hoc frequencies arising likely from the hardware noise picked up from a long-time simulation on the device. 

On the other hand, employing REFF substantially improves the accuracy of the correlation function at long times. Fig.\ref{mit_ex}\textbf{(A)} shows that the observed modulations of the correlation function produced from (indirect) REFF (light orange dots) simulation in the time domain match that of the exact results. 

In the case of a perfect diagonalization, the error remains primarily constant over time since REFF uses a fixed depth circuit and thus does not suffer from an accumulation of hardware errors with simulation time. Prior analysis has shown that REFF suffers an algorithmic error that grows quadratically in time if the diagonalization is imperfect ~\cite{gibbs2022dynamical}. Here we find that in the timescales investigated here ($ 0 \leq Jt \leq 30$, 100 time steps), the REFF correlation function has intact modulations. While in general the diagonalization parameters can be trained on quantum hardware, as in Refs.~\cite{cirstoiu2020variational,gibbs2022dynamical,gibbs2022long}, due to limited hardware access here we use a classically trained ansatz. In the frequency domain, as shown in Fig.\ref{mit_ex}\textbf{(B)}, the REFF simulation results identify the correct peaks in the spectrum. However, we notice that the amplitude of the modulation of the correlation function of REFF is reduced from the expectation from the exact results, resulting in a reduced amplitude of the peaks in the frequency domain. 

 In order to correct the amplitude error in Fig.\ref{mit_ex}\textbf{(A)}-\textbf{(B)}, we explore direct measurements. In Fig.\ref{mit_ex}\textbf{(C)}-\textbf{(D)}, we compare the results of indirect and direct (dark red dots) measurements for Re$(C^{xx}_{1,1}(t))$ using REFF. While the indirect measurements use a standard Hadamard test with controlled operations between an ancillary qubit and system qubits to compute the correlation functions, the direct method uses mid-circuit measurements.  We find that the direct REFF measurements produce near-accurate results shown by the close match of dark red points to the exact results shown in grey lines. The amplitude error incurred by the indirect REFF measurements of Fig.\ref{mit_ex}\textbf{(A)}-\textbf{(B)} are fully mitigated as also shown in the Fourier transform results in Fig.\ref{mit_ex}\textbf{(D)}. This stark improvement highlights that the constant error observed for REFF implemented with indirect measurements is primarily due to the hardware noise induced by the use of the Hadamard test rather than intrinsically from REFF.  It is to be mentioned that the performance of indirect measurements along with the enhancement from switching to direct measurements are dependent on the device used and the calibration of the qubits at run time. Nonetheless, there has always been a degree of improvement using direct measurements over the different sets of results we have obtained. Overall, our results establish REFF simulations implemented with direct measurements as the most desirable approach to a spin-spin correlation measurement on noisy circuit-based quantum hardware. 

 To quantify further the high fidelity of the direct REFF results in Fig.\ref{mit_ex}, in Fig.\ref{err_anal1} we present an error analysis of these results. In \textbf{(A)}, the Root Mean Square (RMS) errors between the measured correlation function and exact values are calculated over time, and in \textbf{(B)}, the corresponding RMS errors of the power spectrum in the frequency domain are presented. We find that the RMS error from direct REFF measurements is very low, having an average value of $\sim 15 \%$.  On careful scrutiny, it is possible to discern that for the very first few steps of time evolution simulation, Trotterization performs better than REFF. This is expected as REFF uses a diagonal \textit{approximation} of the trotter unitary which uses 6 CNOTs compared to a single Trotter step which uses 3 CNOTs. However, this feature is short-lived and by 3-4 time steps, our diagonalized ansatz produces better results as it can avoid accumulating the noise-induced error which comes from the growing circuit depth of Trotterization.

Having developed the implementation of these methods on a noisier backend, it was essential to carefully choose the device/qubits to produce all of the required correlation functions for the dynamical scattering factor calculation, which we describe next.

\medskip
\paragraph*{Dynamical scattering factor calculation:}

To calculate the dynamical scattering factor which is proportional to the intensity spectrum of the system, we calculate the real and imaginary parts of the spin-spin correlation functions $C^{\alpha\beta}_{i,j}(t)$, where $i,j$ represent the indices of the spin sites and $\alpha,\beta = x,y,z$ depending on the chosen polarization of the $\mathbf{Q}$ momentum vector, along with their Fourier transforms. In our case, we chose to simulate for $\mathbf{Q} = 0$, limiting our selection of correlation functions to the case of $\alpha = \beta$. As described in Sec.~\ref{Lehmann}, the correlation functions $C^{xx}_{i,i}(t)=C^{yy}_{i,i}(t)= \frac{1}{2}[e^{-i(4J + 2h)t} + e^{-i(4J - 2h)t}]$ has two peaks in frequency centered at $4J \pm 2h$ while $C^{zz}_{i,i}(t)= [e^{-i(4J)t}]$ has only one peak centered at $4J$. Taken together, these correlation functions represent the famous triplet states of a spin dimer. 

We begin with a careful choice of qubits by investigating the performance of $C^{xx}_{1,1}(t)$ on qubit pairs implementing direct measurements.  We settled on \textit{ibm\_auckland} with the qubits' initial layout of [4,1]. The calibration details of these qubits are mentioned in Table \ref{table:1}. In Fig.\ref{fig5}\textbf{(A)}, we present REFF (dark red dots, pink dots for measurement error mitigated results) and Trotterization (dark blue dots, dark green dots for measurement error mitigated results) results of the real part of the $xx$ correlation function, Re($C^{xx}_{1,1}(t)$) over time, while in Fig.\ref{fig5}\textbf{(B)}, we plot their power spectra in the frequency domain. Measurement error mitigation was also applied. Although this does enhance the accuracy of our results, measurement error mitigation does not show any appreciable improvement. We see a clear advantage of such qubit choice; when compared to Fig.\ref{mit_ex}, the Trotterization results decay much slower (over roughly $\sim 25$ trotter steps, with a half-life of $\sim 10$ steps). The improvement is also apparent in the frequency domain where we could now identify the intensity peaks arising from Trotterization. Not surprisingly, however, REFF summarily produces better results, with a spectrum almost matching the exact results perfectly.

 This is also apparent in  Fig.\ref{err_anal1}\textbf{(C)} and \textbf{(D)}, where we show the corresponding error analysis to these results. In the time domain, the REFF result with direct measurements has an RMS error that has an average value $\sim 5.6\%$, compared to the $\sim$ 40\% error percentage for the result Trotterization result in Fig.\ref{mit_ex}. 
 
In Fig.\ref{fig5}\textbf{(C)}-\textbf{(D)}, we present the result of measuring the real part of the $zz$ correlation function, Re$(C^{zz}_{1,1}(t))$, and its Fourier transform to illustrate the one peak we find at $4J$ in the frequency domain. The complete measurements of the real and imaginary parts of the correlation functions $C^{xx}_{1,1}(t)$, $C^{yy}_{1,1}(t)$ and $C^{zz}_{1,1}(t)$ with their power spectra are illustrated in Fig.\ref{spec_pic}.

From here, we calculate the dynamical structure factor as shown in Eq.\eqref{eq:CorrelationFunc} for the case of $\mathbf{Q} = 0$ for the 2 spin-1/2 system with the Heisenberg interaction ($J = 1$, $h = 1$). This result is presented in Fig.\ref{Heis_Int_Dir}, where we plot the dynamical scattering cross section against frequency $\omega$. The device used was \textit{ibm\_auckland} and the qubits' initial layout was [7,4,1] for the indirect measurements (orange dots, yellow dots for measurement error mitigated results) and [4,1] for the direct measurements (dark red dots, pink dots for the measurement error mitigated results).
In an actual INS experiment, depending on the experimental conditions and scattering geometry, the spectra are weighted differently (see Eq.\eqref{eq:IntensitySpectrum}) when added up. But for the sake of demonstration, here we add up all the spectra isotropically. We clearly can separately identify the three distinct peaks of the triplet states which represent the density of states as measured in a 2-spin dimer in a field, both in the indirect and direct measurement schemes - with the latter showing a near-perfect match with the exact results. The measurement error mitigated results are also provided, however as before, minimal improvement is observed. Our best simulation result, namely using REFF with direct measurements and measurement error mitigation, has an average error of $\sim$ 5.5\%.

\medskip
\paragraph*{General 2-spin models and comparison with experimental data:}
The REFF ansatz is not restricted to the Heisenberg model with a transverse field Hamiltonian where $J_{xx} = J_{yy} = J_{zz}$, as it is possible to re-optimize for more general cases. This is essential since most 2-spin models that manifest in condensed matter or in magnetic molecules end up having some degree of non-Heisenberg anisotropy arising from the local fields. Our ability to accurately calculate the dynamical structure factor for a general spin interaction system allows us to compare simulations to experimental systems - and to experimental data from INS - which could be one enticing futuristic application of this technique. To demonstrate the flexibility of the Hamiltonian we can choose, for our case of the 2-spin Hamiltonian, we take as an example the experimental INS data in Ref.~\cite{kurita2019localized} on the dimerized magnet Ba$_2$CoSi$_2$O$_6$Cl$_2$ for $\mathbf{Q}\sim 0$ (the blue dots in Fig.\ref{sim_Res}). The crystal structure of Ba$_2$CoSi$_2$O$_6$Cl$_2$ is presented in Fig.\ref{fig:dummy}, comprising of a rhombohedral mesh of Copper spin-1/2 atoms (we also present a schematic view of its magnetic structure in Fig.~\ref{fig:dummy}). Under one proposal the excitations could be modelled as a system of isolated dimers governed by an XY model with a ZZ perturbation term: 
\begin{equation}\label{eq:molmodel}
    H = 11.4(X_1 X_2 + Y_1 Y_2) + 0.16(Z_1 Z_2) .
\end{equation}
This model is well-suited for simulation via our REFF ansatz. A detailed discussion on this compound and the INS is presented in Sec.~\ref{compound} as well as in Ref.~\cite{kurita2019localized}.

In Fig.\ref{sim_Res}, we present the results of measuring the intensity spectrum of the dimerized magnet Ba$_2$CoSi$_2$O$_6$Cl$_2$ from INS experiments (teal blue dots) performed in the Materials and Life Science Experimental Facility (MLF) at J-PARC, Japan. The magnetic excitations were investigated via INS experiments using the cold-neutron disk chopper spectrometer AMATERAS \cite{nakajima2011amateras,kurita2019localized} at 4 K with an incident neutron energy of $E_i = 15.2$ meV \cite{kurita2019localized}. The data, integrated in $\mathbf{Q}$ in the range $0.9-1.1 \AA^{-1}$, presented three peaks, although the suggested Hamiltonian only accounts for the middle peak shaded in blue, which we attempt to emulate here. We perform the simulation results of the dynamical scattering factor obtained from direct REFF measurements (dark red dots, pink dots for measurement error mitigated results) on quantum hardware as a function of energy. The data obtained was on powder. Hence, a simple summation of the dimer structure factor with isotropically added coefficients is a good approximation to a powder average for the energy density of states. We use the same procedure as in Fig.\ref{Heis_Int_Dir} -  we calculate the dynamical scattering cross-section compiled from the correlation function obtained using REFF, direct measurements, and measurement error mitigation, for which these results were run on \textit{ibm\_hanoi} with the qubits' initial layout as [9,8]. The calibration details of the qubits are mentioned in Table \ref{table:3}. 
 
As before, the grey lines are the exact result for Eq.\eqref{eq:molmodel}. We find that the average error between the intensity spectrum predicted by the hardware results and the exact value for the perturbed XY model, Eq.\eqref{eq:molmodel}, is less than 1\%. 
The exact momentum dependence of the data would need a full simulation of the exact geometrical considerations of the experiment and proper accounting of the form factors and $k/k'$ ratios, but here we compare only the energy dependence. We find that the simulation, with $\sim$ 30 trotter steps, is sufficient to capture the energy modes with the requisite resolution better than the INS data (taken at $E_i = 15.2$ meV, which is quite typical of neutron energies delivered by standard cold and sub-thermal neutron sources - both reactor or spallation types - around the world \cite{adams2020first}). While for a more general higher-spin Hamiltonian, we expect more quadratic noise, the comparison would suffice if REFF optimization accurately captures the evolution in the first few tens of iterations. 

\section{Discussion}

Simulating dynamics and interpreting spectroscopic measurements for quantum many-body systems is generally challenging for classical computers \cite{georgescu2014quantum}.  When studying the dynamics of a system by simulating its temporal evolution, these simulations require a number of operations that increases exponentially with the size of the system \cite{georgescu2014quantum, chiesa2019quantum}, demanding impractically large computational memory and/or execution time. Moreover, quantum Monte Carlo methods face great limitations when studying more complex systems, such as the sign problem \cite{troyer2005computational, pan2022sign}. For quantum computers, however, such simulations are of bounded-error quantum polynomial time (BQP) complexity \cite{feynman1985quantum, nagaj2010fast, muller2015promise}. Hence, quantum computers are predicted to exceed classical computers in their capabilities in the simulation of quantum systems \cite{shor1999polynomial,farhi2014quantum,feynman2018simulating}. 

As presented in this paper, the use of quantum simulation opens new avenues for simulating inelastic neutron scattering experiments more efficiently. We obtained quantum simulation results for the correlation functions at temperature $T = 0$ of 2 spin-1/2 linear chains (isolated dimers) with Heisenberg interactions between the spins, from which we were able to calculate the dynamical structure factor. We found that the use of REFF for time evolution simulation, direct measurements, and measurement error mitigation all give us results of greater accuracy and more noise resilience as compared to the more standard quantum simulation methods.  

We have shown that the simulations performed in this fashion can significantly improve solution quality to reveal robust spin gaps and splittings between levels in magnetic models. The results when computed over $~30$ time evolution steps can be Fourier transformed to reveal the energy spectrum of the dynamical structure factor with sufficient resolution to start matching to current experimental data obtained in quantum magnets. 

While our first demonstration in this manuscript is on a 2-spin system, we expect that the methodology presented here can be extended to many-body Hamiltonians. REFF approximately diagonalizes the Hamiltonian, allowing long-time simulations in significantly shorter depth circuits. As we had mentioned earlier, the accuracy of REFF has a quadratic dropoff with evolution time. However, if comparison to INS data is our objective, this accuracy needs to be only sufficiently good to allow for ~30 - 50 time steps without dephasing errors overwhelming the correlation coefficients. Sufficient bandwidth for a meaningful Fourier analysis to match the data/resolutions achieved in INS experiments could be achieved in this way.  

Attempts at measuring the correlation functions at $T = 0$ have been presented in other papers. For example, in terms of the time evolution simulation, Ref.~\cite{chiesa2019quantum} used Trotterization to approximate the time evolution operator and Ref.~\cite{francis2020quantum} was able to make use of the special properties of their chosen Hamiltonian model to implement a Cartan KAK decomposition~\cite{vidal2004universal}. Both implemented indirect measurements. 

In terms of scaling up to truly many-body spin systems (such as spanning dozens of spins), more research is necessary. This is because REFF requires an optimization process that would become much more complicated the larger the simulated system. However, REFF has been proven to be much more noise-resilient in comparison to Trotterization. Hence, we aim to further study more efficient methods for the diagonalization of Hamiltonians and fast-forwarding of time evolution, along with studying lattices of dimers which could hopefully lead us to simulate emergent quantum behaviour such as those observed in RVB states. There are already attempts at implementing different variational algorithms to facilitate the diagonalization process, such as diagonalizing over the basis of a specific initial state instead of the whole Hilbert space as described in Refs.\cite{lim2021fast, gibbs2022dynamical}. This scaling-up process will additionally necessitate that we implement further error mitigation techniques and have a deeper understanding of the noise in the hardware affecting our results. Hence, these will be points of immense interest as we progress with our work, aiming to simulate larger and more complex systems.

As we have previously mentioned, dimers are the main building blocks of exotic quantum states like RVBs and QSLs \cite{broholm2020quantum, haravifard2014emergence}. They are also building blocks for spin-dimer qubits \cite{bartling2022entanglement} such as in $C^{13}$ dimers in diamonds where long-timescale dephasing of a dimer and inter-dimer interactions could be cross-benchmarked. Hence, the entanglement of several dimers is an excellent scaling-up route to go.

Another challenge of this field is the initial state preparation stage of the algorithm will become more challenging for many-body Hamiltonians. Many Hamiltonians have ground states that cannot be analytically obtained, and hence, we would need to use state preparation algorithms. In this case, one could consider using the Variational Quantum Eigensolver (VQE) algorithm \cite{Qiskit, tilly2022variational} or the Quantum Approximate Optimizer Algorithm (QAOA)\cite{Qiskit, farhi2014quantum}. Also of prime importance will be methods such as classical Metropolis or Luttinger-Tisza methods \cite{chib1995understanding,robert2010metropolis,luttinger1946theory,litvin1974luttinger} to reveal insights into efficient preparation of the ground state for time-evolution. 

To get a more accurate representation of the actual dynamics of physical systems one could extend the algorithm to compute finite temperature correlation functions. While the literature on preparing thermal states on quantum hardware is extensive~\cite{brandao2019finite,cohn2020minimal,sun2021quantum}, implementations typically require deep circuits, prior knowledge of the system's spectral properties or are non-scalable. Thus more improvements in hardware and/or software are needed to push our work in this direction.

The future relationship between quantum simulation and INS experiments is likely to be multifaceted. In the near term, we predominantly see quantum simulations being used for \textit{benchmarking}. That is, quantum simulations, INS experiments and classical simulations of the same systems can be performed and compared to benchmark each another. Longer-term, once the accuracy of quantum simulations has been established, quantum simulations could be used for \textit{prediction}, thereby aiding the more costly and time-consuming INS experiments on carefully grown large samples. We believe that in the future simulations of INS outputs could even be combined with meta-learning strategies to search for new materials. Or the outputs of INS experiments could be combined with Hamiltonian learning strategies and quantum simulation to search for models that explain INS outputs. 
In all cases, we expect improved methods for quantum simulation, of the sort studied here, will prove critical.

\newpage

\bibliographystyle{ieeetr}

\begin{thebibliography}{}

\end{thebibliography}


\begin{thebibliography}{10}

\bibitem{lupke2022local}
F.~L{\"u}pke, A.~D. Pham, Y.-F. Zhao, L.-J. Zhou, W.~Lu, E.~Briggs,
  J.~Bernholc, M.~Kolmer, J.~Teeter, W.~Ko, {\em et~al.}, ``{L}ocal
  manifestations of thickness-dependent topology and edge states in the
  topological magnet {Mn}{Bi}$_2${Te}$_4$,'' {\em Physical Review B}, vol.~105,
  no.~3, p.~035423, 2022.

\bibitem{hasan2010colloquium}
M.~Z. Hasan and C.~L. Kane, ``{C}olloquium: {{T}opological} insulators,'' {\em
  {Reviews of Modern Physics}}, vol.~82, no.~4, p.~3045, 2010.

\bibitem{he2019topological}
M.~He, H.~Sun, and Q.~L. He, ``Topological insulator: Spintronics and quantum
  computations,'' {\em Frontiers of Physics}, vol.~14, no.~4, pp.~1--16, 2019.

\bibitem{bi2022drastic}
W.~Bi, T.~Culverhouse, Z.~Nix, W.~Xie, H.-J. Tien, T.-R. Chang, U.~Dutta,
  J.~Zhao, B.~Lavina, E.~E. Alp, {\em et~al.}, ``{D}rastic enhancement of
  magnetic critical temperature and amorphization in topological magnet
  \ce{EuSn$_2$P$_2$} under pressure,'' {\em npj Quantum Materials}, vol.~7,
  no.~1, pp.~1--8, 2022.

\bibitem{serrano2022magnetic}
G.~Serrano, L.~Poggini, G.~Cucinotta, A.~Sorrentino, N.~Giaconi, B.~Cortigiani,
  D.~Longo, E.~Otero, P.~Sainctavit, A.~Caneschi, {\em et~al.}, ``Magnetic
  molecules as sensors of topological hysteresis of superconductors,'' 2022.

\bibitem{chiesa2019quantum}
A.~Chiesa, F.~Tacchino, M.~Grossi, P.~Santini, I.~Tavernelli, D.~Gerace, and
  S.~Carretta, ``Quantum hardware simulating four-dimensional inelastic neutron
  scattering,'' {\em Nature Physics}, vol.~15, no.~5, pp.~455--459, 2019.

\bibitem{moreno2021measuring}
E.~Moreno-Pineda and W.~Wernsdorfer, ``Measuring molecular magnets for quantum
  technologies,'' {\em Nature Reviews Physics}, vol.~3, no.~9, pp.~645--659,
  2021.

\bibitem{anderson1973resonating}
P.~W. Anderson, ``Resonating valence bonds: A new kind of insulator?,'' {\em
  Materials Research Bulletin}, vol.~8, no.~2, pp.~153--160, 1973.

\bibitem{balents2010spin}
L.~Balents, ``Spin liquids in frustrated magnets,'' {\em Nature}, vol.~464,
  no.~7286, pp.~199--208, 2010.

\bibitem{ebadi2021quantum}
S.~Ebadi, T.~T. Wang, H.~Levine, A.~Keesling, G.~Semeghini, A.~Omran,
  D.~Bluvstein, R.~Samajdar, H.~Pichler, W.~W. Ho, {\em et~al.}, ``Quantum
  phases of matter on a 256-atom programmable quantum simulator,'' {\em
  Nature}, vol.~595, no.~7866, pp.~227--232, 2021.

\bibitem{richter2021simulating}
J.~Richter and A.~Pal, ``Simulating hydrodynamics on noisy intermediate-scale
  quantum devices with random circuits,'' {\em Physical Review Letters},
  vol.~126, no.~23, p.~230501, 2021.

\bibitem{kalinowski2022non}
M.~Kalinowski, N.~Maskara, and M.~D. Lukin, ``Non-abelian floquet spin liquids
  in a digital rydberg simulator,'' {\em arXiv preprint arXiv:2211.00017},
  2022.

\bibitem{banerjee2017neutron}
A.~Banerjee, J.~Yan, J.~Knolle, C.~A. Bridges, M.~B. Stone, M.~D. Lumsden,
  D.~G. Mandrus, D.~A. Tennant, R.~Moessner, and S.~E. Nagler, ``{N}eutron
  scattering in the proximate quantum spin liquid $\alpha$-\ce{RuCl$_3$},''
  {\em Science}, vol.~356, no.~6342, pp.~1055--1059, 2017.

\bibitem{anderson1993resonating}
P.~Anderson, ``{T}he resonating valence bond state in \ce{La$_2$CuO$_4$} and
  superconductivity,'' in {\em Ten Years of Superconductivity: 1980--1990},
  pp.~278--280, Springer, 1993.

\bibitem{urushihara2020crystal}
D.~Urushihara, S.~Kawaguchi, K.~Fukuda, and T.~Asaka, ``{C}rystal structure and
  magnetism in the {S} = 1/2 spin dimer compound
  \ce{{Na}{Cu}$_2${VP}$_2${O}$_{10}$},'' {\em IUCrJ}, vol.~7, no.~4,
  pp.~656--662, 2020.

\bibitem{broholm2020quantum}
C.~Broholm, R.~Cava, S.~Kivelson, D.~Nocera, M.~Norman, and T.~Senthil,
  ``Quantum spin liquids,'' {\em Science}, vol.~367, no.~6475, p.~eaay0668,
  2020.

\bibitem{chaloupka2010kitaev}
J.~Chaloupka, G.~Jackeli, and G.~Khaliullin, ``{K}itaev-{H}eisenberg model on a
  honeycomb lattice: possible exotic phases in iridium oxides
  \ce{{A}$_2${Ir}{O}$_3$},'' {\em {P}hysical {R}eview {L}etters}, vol.~105,
  no.~2, p.~027204, 2010.

\bibitem{veiga2017pressure}
L.~Veiga, M.~Etter, K.~Glazyrin, F.~Sun, C.~Escanhoela~Jr, G.~Fabbris,
  J.~Mardegan, P.~Malavi, Y.~Deng, P.~Stavropoulos, {\em et~al.}, ``Pressure
  tuning of bond-directional exchange interactions and magnetic frustration in
  the hyperhoneycomb iridate $\beta$-\ce{{Li}$_2${Ir}{O}$_3$},'' {\em Physical
  Review B}, vol.~96, no.~14, p.~140402, 2017.

\bibitem{plumb2014alpha}
K.~Plumb, J.~Clancy, L.~Sandilands, V.~V. Shankar, Y.~Hu, K.~Burch, H.-Y. Kee,
  and Y.-J. Kim, ``$\alpha$-\ce{{Ru}{Cl}$_3$}: A spin-orbit assisted mott
  insulator on a honeycomb lattice,'' {\em Physical Review B}, vol.~90, no.~4,
  p.~041112, 2014.

\bibitem{takagi2019concept}
H.~Takagi, T.~Takayama, G.~Jackeli, G.~Khaliullin, and S.~E. Nagler,
  ``{C}oncept and realization of {{K}itaev} quantum spin liquids,'' {\em Nature
  Reviews Physics}, vol.~1, no.~4, pp.~264--280, 2019.

\bibitem{banerjee2016proximate}
A.~Banerjee, C.~Bridges, J.-Q. Yan, A.~Aczel, L.~Li, M.~Stone, G.~Granroth,
  M.~Lumsden, Y.~Yiu, J.~Knolle, {\em et~al.}, ``{P}roximate {{K}itaev} quantum
  spin liquid behaviour in a honeycomb magnet,'' {\em Nature materials},
  vol.~15, no.~7, pp.~733--740, 2016.

\bibitem{haravifard2014emergence}
S.~Haravifard, A.~Banerjee, J.~van Wezel, D.~Silevitch, A.~dos Santos, J.~Lang,
  E.~Kermarrec, G.~Srajer, B.~D. Gaulin, J.~Molaison, {\em et~al.}, ``Emergence
  of long-range order in sheets of magnetic dimers,'' {\em Proceedings of the
  National Academy of Sciences}, vol.~111, no.~40, pp.~14372--14377, 2014.

\bibitem{furrer2009neutron}
A.~Furrer, J.~F. Mesot, and T.~Str{\"a}ssle, {\em Neutron scattering in
  condensed matter physics}, vol.~4.
\newblock World Scientific Publishing Company, 2009.

\bibitem{adams2020first}
P.~Adams, J.~F. Ankner, L.-L. Anovitz, A.~Banerjee, E.~Begoli, R.~Boehler,
  S.~Calder, B.~C. Chakoumakos, T.~R. Charlton, W.-R. Chen, {\em et~al.},
  ``First experiments: New science opportunities at the spallation neutron
  source second target station (abridged),'' tech. rep., Oak Ridge National
  Lab.(ORNL), Oak Ridge, TN (United States), 2020.

\bibitem{ortiz2001quantum}
G.~Ortiz, J.~E. Gubernatis, E.~Knill, and R.~Laflamme, ``Quantum algorithms for
  fermionic simulations,'' {\em Physical Review A}, vol.~64, no.~2, p.~022319,
  2001.

\bibitem{bravyi2002fermionic}
S.~B. Bravyi and A.~Y. Kitaev, ``Fermionic quantum computation,'' {\em Annals
  of Physics}, vol.~298, no.~1, pp.~210--226, 2002.

\bibitem{huggins2022unbiasing}
W.~J. Huggins, B.~A. O’Gorman, N.~C. Rubin, D.~R. Reichman, R.~Babbush, and
  J.~Lee, ``Unbiasing fermionic quantum monte carlo with a quantum computer,''
  {\em Nature}, vol.~603, no.~7901, pp.~416--420, 2022.

\bibitem{bartling2022entanglement}
H.~Bartling, M.~Abobeih, B.~Pingault, M.~Degen, S.~Loenen, C.~Bradley,
  J.~Randall, M.~Markham, D.~Twitchen, and T.~Taminiau, ``Entanglement of
  spin-pair qubits with intrinsic dephasing times exceeding a minute,'' {\em
  Physical Review X}, vol.~12, no.~1, p.~011048, 2022.

\bibitem{lovesey1984theory}
S.~W. Lovesey, ``Theory of neutron scattering from condensed matter,'' 1984.

\bibitem{suzuki1993improved}
M.~Suzuki, ``Improved trotter-like formula,'' {\em Physics Letters A},
  vol.~180, no.~3, pp.~232--234, 1993.

\bibitem{sornborger1999higher}
A.~Sornborger, ``Higher-order methods for simulations on quantum computers,''
  {\em Physical Review A}, vol.~60, no.~3, p.~1956, 1999.

\bibitem{gibbs2022dynamical}
J.~Gibbs, Z.~Holmes, M.~C. Caro, N.~Ezzell, H.-Y. Huang, L.~Cincio, A.~T.
  Sornborger, and P.~J. Coles, ``Dynamical simulation via quantum machine
  learning with provable generalization,'' {\em arXiv preprint
  arXiv:2204.10269}, 2022.

\bibitem{knee2015optimal}
G.~C. Knee and W.~J. Munro, ``Optimal trotterization in universal quantum
  simulators under faulty control,'' {\em Physical Review A}, vol.~91, no.~5,
  p.~052327, 2015.

\bibitem{cirstoiu2020variational}
C.~Cirstoiu, Z.~Holmes, J.~Iosue, L.~Cincio, P.~J. Coles, and A.~Sornborger,
  ``Variational fast forwarding for quantum simulation beyond the coherence
  time,'' {\em npj Quantum Information}, vol.~6, no.~1, pp.~1--10, 2020.

\bibitem{commeau2020variational}
B.~Commeau, M.~Cerezo, Z.~Holmes, L.~Cincio, P.~J. Coles, and A.~Sornborger,
  ``Variational {H}amiltonian diagonalization for dynamical quantum
  simulation,'' {\em arXiv preprint arXiv:2009.02559}, 2020.

\bibitem{gibbs2022long}
J.~Gibbs, K.~Gili, Z.~Holmes, B.~Commeau, A.~Arrasmith, L.~Cincio, P.~J. Coles,
  and A.~Sornborger, ``Long-time simulations for fixed input states on quantum
  hardware,'' {\em npj Quantum Information}, vol.~8, no.~1, p.~135, 2022.

\bibitem{geller2021experimental}
M.~R. Geller, Z.~Holmes, P.~J. Coles, and A.~Sornborger, ``Experimental quantum
  learning of a spectral decomposition,'' {\em Physical Review Research},
  vol.~3, no.~3, p.~033200, 2021.

\bibitem{caro2022generalization}
M.~C. Caro, H.-Y. Huang, M.~Cerezo, K.~Sharma, A.~Sornborger, L.~Cincio, and
  P.~J. Coles, ``Generalization in quantum machine learning from few training
  data,'' {\em Nature communications}, vol.~13, no.~1, p.~4919, 2022.

\bibitem{caro2022outofdistribution}
M.~C. Caro, H.-Y. Huang, N.~Ezzell, J.~Gibbs, A.~T. Sornborger, L.~Cincio,
  P.~J. Coles, and Z.~Holmes, ``Out-of-distribution generalization for learning
  quantum dynamics,'' {\em arXiv preprint arXiv:2204.10268}, 2022.

\bibitem{somma2002simulating}
R.~Somma, G.~Ortiz, J.~E. Gubernatis, E.~Knill, and R.~Laflamme, ``Simulating
  physical phenomena by quantum networks,'' {\em Physical Review A}, vol.~65,
  no.~4, p.~042323, 2002.

\bibitem{mitarai2019methodology}
K.~Mitarai and K.~Fujii, ``Methodology for replacing indirect measurements with
  direct measurements,'' {\em Physical Review Research}, vol.~1, no.~1,
  p.~013006, 2019.

\bibitem{maciejewski2020mitigation}
F.~B. Maciejewski, Z.~Zimbor{\'a}s, and M.~Oszmaniec, ``Mitigation of readout
  noise in near-term quantum devices by classical post-processing based on
  detector tomography,'' {\em Quantum}, vol.~4, p.~257, 2020.

\bibitem{Qiskit}
M.~S. ANIS, Abby-Mitchell, H.~Abraham, AduOffei, R.~Agarwal, G.~Agliardi,
  M.~Aharoni, V.~Ajith, I.~Y. Akhalwaya, G.~Aleksandrowicz, T.~Alexander,
  M.~Amy, S.~Anagolum, Anthony-Gandon, I.~F. Araujo, E.~Arbel, A.~Asfaw,
  A.~Athalye, A.~Avkhadiev, C.~Azaustre, P.~BHOLE, A.~Banerjee, S.~Banerjee,
  W.~Bang, A.~Bansal, P.~Barkoutsos, A.~Barnawal, G.~Barron, G.~S. Barron,
  L.~Bello, Y.~Ben-Haim, M.~C. Bennett, D.~Bevenius, D.~Bhatnagar,
  P.~Bhatnagar, A.~Bhobe, P.~Bianchini, L.~S. Bishop, C.~Blank, S.~Bolos,
  S.~Bopardikar, S.~Bosch, S.~Brandhofer, Brandon, S.~Bravyi, N.~Bronn,
  Bryce-Fuller, D.~Bucher, A.~Burov, F.~Cabrera, P.~Calpin, L.~Capelluto,
  J.~Carballo, G.~Carrascal, A.~Carriker, I.~Carvalho, A.~Chen, C.-F. Chen,
  E.~Chen, J.~C. Chen, R.~Chen, F.~Chevallier, K.~Chinda, R.~Cholarajan, J.~M.
  Chow, S.~Churchill, CisterMoke, C.~Claus, C.~Clauss, C.~Clothier, R.~Cocking,
  R.~Cocuzzo, J.~Connor, F.~Correa, Z.~Crockett, A.~J. Cross, A.~W. Cross,
  S.~Cross, J.~Cruz-Benito, C.~Culver, A.~D. C{\'o}rcoles-Gonzales, N.~D,
  S.~Dague, T.~E. Dandachi, A.~N. Dangwal, J.~Daniel, M.~Daniels, M.~Dartiailh,
  A.~R. Davila, F.~Debouni, A.~Dekusar, A.~Deshmukh, M.~Deshpande, D.~Ding,
  J.~Doi, E.~M. Dow, P.~Downing, E.~Drechsler, E.~Dumitrescu, K.~Dumon,
  I.~Duran, K.~EL-Safty, E.~Eastman, G.~Eberle, A.~Ebrahimi, P.~Eendebak,
  D.~Egger, ElePT, Emilio, A.~Espiricueta, M.~Everitt, D.~Facoetti, Farida,
  P.~M. Fern{\'a}ndez, S.~Ferracin, D.~Ferrari, A.~H. Ferrera, R.~Fouilland,
  A.~Frisch, A.~Fuhrer, B.~Fuller, M.~GEORGE, J.~Gacon, B.~G. Gago,
  C.~Gambella, J.~M. Gambetta, A.~Gammanpila, L.~Garcia, T.~Garg, S.~Garion,
  J.~R. Garrison, J.~Garrison, T.~Gates, H.~Georgiev, L.~Gil, A.~Gilliam,
  A.~Giridharan, Glen, J.~Gomez-Mosquera, Gonzalo, S.~de~la
  Puente~Gonz{\'a}lez, J.~Gorzinski, I.~Gould, D.~Greenberg, D.~Grinko,
  W.~Guan, D.~Guijo, Guillermo-Mijares-Vilarino, J.~A. Gunnels, H.~Gupta,
  N.~Gupta, J.~M. G{\"u}nther, M.~Haglund, I.~Haide, I.~Hamamura, O.~C. Hamido,
  F.~Harkins, K.~Hartman, A.~Hasan, V.~Havlicek, J.~Hellmers, {\L}.~Herok,
  S.~Hillmich, C.~Hong, H.~Horii, C.~Howington, S.~Hu, W.~Hu, C.-H. Huang,
  J.~Huang, R.~Huisman, H.~Imai, T.~Imamichi, K.~Ishizaki, Ishwor, R.~Iten,
  T.~Itoko, A.~Ivrii, A.~Javadi, A.~Javadi-Abhari, W.~Javed, Q.~Jianhua,
  M.~Jivrajani, K.~Johns, S.~Johnstun, Jonathan-Shoemaker, JosDenmark,
  JoshDumo, J.~Judge, T.~Kachmann, A.~Kale, N.~Kanazawa, J.~Kane, Kang-Bae,
  A.~Kapila, A.~Karazeev, P.~Kassebaum, T.~Kehrer, J.~Kelso, S.~Kelso, H.~van
  Kemenade, V.~Khanderao, S.~King, Y.~Kobayashi, Kovi11Day, A.~Kovyrshin,
  R.~Krishnakumar, P.~Krishnamurthy, V.~Krishnan, K.~Krsulich, P.~Kumkar,
  G.~Kus, R.~LaRose, E.~Lacal, R.~Lambert, H.~Landa, J.~Lapeyre, J.~Latone,
  S.~Lawrence, C.~Lee, G.~Li, T.~J. Liang, J.~Lishman, D.~Liu, P.~Liu, Lolcroc,
  A.~K. M, L.~Madden, Y.~Maeng, S.~Maheshkar, K.~Majmudar, A.~Malyshev, M.~E.
  Mandouh, J.~Manela, Manjula, J.~Marecek, M.~Marques, K.~Marwaha, D.~Maslov,
  P.~Maszota, D.~Mathews, A.~Matsuo, F.~Mazhandu, D.~McClure, M.~McElaney,
  J.~McElroy, C.~McGarry, D.~McKay, D.~McPherson, S.~Meesala, D.~Meirom,
  C.~Mendell, T.~Metcalfe, M.~Mevissen, A.~Meyer, A.~Mezzacapo, R.~Midha,
  D.~Miller, H.~Miller, Z.~Minev, A.~Mitchell, N.~Moll, A.~Montanez,
  G.~Monteiro, M.~D. Mooring, R.~Morales, N.~Moran, D.~Morcuende, S.~Mostafa,
  M.~Motta, R.~Moyard, P.~Murali, D.~Murata, J.~M{\"u}ggenburg, T.~NEMOZ,
  D.~Nadlinger, K.~Nakanishi, G.~Nannicini, P.~Nation, E.~Navarro, Y.~Naveh,
  S.~W. Neagle, P.~Neuweiler, A.~Ngoueya, T.~Nguyen, J.~Nicander,
  Nick-Singstock, P.~Niroula, H.~Norlen, NuoWenLei, L.~J. O'Riordan,
  O.~Ogunbayo, P.~Ollitrault, T.~Onodera, R.~Otaolea, S.~Oud, D.~Padilha,
  H.~Paik, S.~Pal, Y.~Pang, A.~Panigrahi, V.~R. Pascuzzi, S.~Perriello,
  E.~Peterson, A.~Phan, K.~Pilch, F.~Piro, M.~Pistoia, C.~Piveteau, J.~Plewa,
  P.~Pocreau, A.~Pozas-Kerstjens, R.~Pracht, M.~Prokop, V.~Prutyanov, S.~Puri,
  D.~Puzzuoli, Pythonix, J.~P{\'e}rez, Quant02, Quintiii, R.~I. Rahman,
  A.~Raja, R.~Rajeev, I.~Rajput, N.~Ramagiri, A.~Rao, R.~Raymond,
  O.~Reardon-Smith, R.~M.-C. Redondo, M.~Reuter, J.~Rice, M.~Riedemann,
  Rietesh, D.~Risinger, P.~Rivero, M.~L. Rocca, D.~M. Rodr{\'\i}guez,
  RohithKarur, B.~Rosand, M.~Rossmannek, M.~Ryu, T.~SAPV, N.~R.~C. Sa, A.~Saha,
  A.~Ash-Saki, S.~Sanand, M.~Sandberg, H.~Sandesara, R.~Sapra, H.~Sargsyan,
  A.~Sarkar, N.~Sathaye, N.~Savola, B.~Schmitt, C.~Schnabel, Z.~Schoenfeld,
  T.~L. Scholten, E.~Schoute, M.~Schulterbrandt, J.~Schwarm, J.~Seaward, Sergi,
  I.~F. Sertage, K.~Setia, F.~Shah, N.~Shammah, W.~Shanks, R.~Sharma, P.~Shaw,
  Y.~Shi, J.~Shoemaker, A.~Silva, A.~Simonetto, D.~Singh, D.~Singh, P.~Singh,
  P.~Singkanipa, Y.~Siraichi, Siri, J.~Sistos, I.~Sitdikov, S.~Sivarajah,
  Slavikmew, M.~B. Sletfjerding, J.~A. Smolin, M.~Soeken, I.~O. Sokolov,
  I.~Sokolov, V.~P. Soloviev, SooluThomas, Starfish, D.~Steenken,
  M.~Stypulkoski, A.~Suau, S.~Sun, K.~J. Sung, M.~Suwama, O.~S{\l}owik,
  R.~Taeja, H.~Takahashi, T.~Takawale, I.~Tavernelli, C.~Taylor, P.~Taylour,
  S.~Thomas, K.~Tian, M.~Tillet, M.~Tod, M.~Tomasik, C.~Tornow, E.~de~la Torre,
  J.~L.~S. Toural, K.~Trabing, M.~Treinish, D.~Trenev, TrishaPe, F.~Truger,
  G.~Tsilimigkounakis, D.~Tulsi, D.~Tuna, W.~Turner, Y.~Vaknin, C.~R. Valcarce,
  F.~Varchon, A.~Vartak, A.~C. Vazquez, P.~Vijaywargiya, V.~Villar, B.~Vishnu,
  D.~Vogt-Lee, C.~Vuillot, J.~Weaver, J.~Weidenfeller, R.~Wieczorek, J.~A.
  Wildstrom, J.~Wilson, E.~Winston, WinterSoldier, J.~J. Woehr, S.~Woerner,
  R.~Woo, C.~J. Wood, R.~Wood, S.~Wood, J.~Wootton, M.~Wright, L.~Xing, J.~YU,
  Yaiza, B.~Yang, U.~Yang, J.~Yao, D.~Yeralin, R.~Yonekura, D.~Yonge-Mallo,
  R.~Yoshida, R.~Young, J.~Yu, L.~Yu, Yuma-Nakamura, C.~Zachow, L.~Zdanski,
  H.~Zhang, I.~Zidaru, B.~Zimmermann, C.~Zoufal, aeddins ibm, alexzhang13, b63,
  bartek bartlomiej, bcamorrison, brandhsn, chetmurthy, choerst ibm,
  deeplokhande, dekel.meirom, dime10, dlasecki, ehchen, ewinston, fanizzamarco,
  fs1132429, gadial, galeinston, georgezhou20, georgios ts, gruu, hhorii,
  hhyap, hykavitha, itoko, jeppevinkel, jessica angel7, jezerjojo14, jliu45,
  johannesgreiner, jscott2, kUmezawa, klinvill, krutik2966, ma5x, michelle4654,
  msuwama, nico lgrs, nrhawkins, ntgiwsvp, ordmoj, sagar pahwa,
  pritamsinha2304, rithikaadiga, ryancocuzzo, saktar unr, saswati qiskit,
  septembrr, sethmerkel, sg495, shaashwat, smturro2, sternparky, strickroman,
  tigerjack, tsura crisaldo, upsideon, vadebayo49, welien, willhbang, wmurphy
  collabstar, yang.luh, yuri@FreeBSD, and M.~{\v{C}}epulkovskis, ``Qiskit: An
  open-source framework for quantum computing,'' 2021.

\bibitem{kurita2019localized}
N.~Kurita, D.~Yamamoto, T.~Kanesaka, N.~Furukawa, S.~Ohira-Kawamura,
  K.~Nakajima, and H.~Tanaka, ``{L}ocalized magnetic excitations in the fully
  frustrated dimerized magnet \ce{{Ba}$_2${Co}{Si}$_2${O}$_6${Cl}$_2$},'' {\em
  Physical Review Letters}, vol.~123, no.~2, p.~027206, 2019.

\bibitem{nakajima2011amateras}
K.~Nakajima, S.~Ohira-Kawamura, T.~Kikuchi, M.~Nakamura, R.~Kajimoto,
  Y.~Inamura, N.~Takahashi, K.~Aizawa, K.~Suzuya, K.~Shibata, {\em et~al.},
  ``Amateras: a cold-neutron disk chopper spectrometer,'' {\em Journal of the
  Physical Society of Japan}, vol.~80, no.~Suppl. B, p.~SB028, 2011.

\bibitem{georgescu2014quantum}
I.~M. Georgescu, S.~Ashhab, and F.~Nori, ``Quantum simulation,'' {\em Reviews
  of Modern Physics}, vol.~86, no.~1, p.~153, 2014.

\bibitem{troyer2005computational}
M.~Troyer and U.-J. Wiese, ``Computational complexity and fundamental
  limitations to fermionic quantum monte carlo simulations,'' {\em {P}hysical
  {R}eview {L}etters}, vol.~94, no.~17, p.~170201, 2005.

\bibitem{pan2022sign}
G.~Pan and Z.~Y. Meng, ``Sign problem in quantum monte carlo simulation,'' {\em
  arXiv preprint arXiv:2204.08777}, 2022.

\bibitem{feynman1985quantum}
R.~P. Feynman, ``Quantum mechanical computers,'' {\em Optics news}, vol.~11,
  no.~2, pp.~11--20, 1985.

\bibitem{nagaj2010fast}
D.~Nagaj, ``{F}ast universal quantum computation with railroad-switch local
  {H}amiltonians,'' {\em Journal of Mathematical Physics}, vol.~51, no.~6,
  p.~062201, 2010.

\bibitem{muller2015promise}
R.~P. Muller and R.~Blume-Kohout, ``The promise of quantum simulation,'' {\em
  ACS nano}, vol.~9, no.~8, pp.~7738--7741, 2015.

\bibitem{shor1999polynomial}
P.~W. Shor, ``Polynomial-time algorithms for prime factorization and discrete
  logarithms on a quantum computer,'' {\em SIAM review}, vol.~41, no.~2,
  pp.~303--332, 1999.

\bibitem{farhi2014quantum}
E.~Farhi, J.~Goldstone, and S.~Gutmann, ``A quantum approximate optimization
  algorithm,'' {\em arXiv preprint arXiv:1411.4028}, 2014.

\bibitem{feynman2018simulating}
R.~P. Feynman, ``Simulating physics with computers,'' in {\em Feynman and
  computation}, pp.~133--153, CRC Press, 2018.

\bibitem{francis2020quantum}
A.~Francis, J.~Freericks, and A.~Kemper, ``Quantum computation of magnon
  spectra,'' {\em Physical Review B}, vol.~101, no.~1, p.~014411, 2020.

\bibitem{vidal2004universal}
G.~Vidal and C.~M. Dawson, ``Universal quantum circuit for two-qubit
  transformations with three controlled-not gates,'' {\em Physical Review A},
  vol.~69, no.~1, p.~010301, 2004.

\bibitem{lim2021fast}
K.~H. Lim, T.~Haug, L.~C. Kwek, and K.~Bharti, ``{F}ast-forwarding with {NISQ}
  processors without feedback loop,'' {\em Quantum Science and Technology},
  vol.~7, no.~1, p.~015001, 2021.

\bibitem{tilly2022variational}
J.~Tilly, H.~Chen, S.~Cao, D.~Picozzi, K.~Setia, Y.~Li, E.~Grant, L.~Wossnig,
  I.~Rungger, G.~H. Booth, {\em et~al.}, ``The variational quantum eigensolver:
  a review of methods and best practices,'' {\em Physics Reports}, vol.~986,
  pp.~1--128, 2022.

\bibitem{chib1995understanding}
S.~Chib and E.~Greenberg, ``Understanding the {M}etropolis-{H}astings
  algorithm,'' {\em {T}he {A}merican {S}tatistician}, vol.~49, no.~4,
  pp.~327--335, 1995.

\bibitem{robert2010metropolis}
C.~Robert, G.~Casella, C.~P. Robert, and G.~Casella, ``Metropolis-{H}astings
  algorithms,'' {\em Introducing Monte Carlo Methods with R}, pp.~167--197,
  2010.

\bibitem{luttinger1946theory}
J.~Luttinger and L.~Tisza, ``Theory of dipole interaction in crystals,'' {\em
  Physical Review}, vol.~70, no.~11-12, p.~954, 1946.

\bibitem{litvin1974luttinger}
D.~B. Litvin, ``The {L}uttinger-{T}isza method,'' {\em Physica}, vol.~77,
  no.~2, pp.~205--219, 1974.

\bibitem{brandao2019finite}
F.~G. Brand{\~a}o and M.~J. Kastoryano, ``Finite correlation length implies
  efficient preparation of quantum thermal states,'' {\em Communications in
  Mathematical Physics}, vol.~365, pp.~1--16, 2019.

\bibitem{cohn2020minimal}
J.~Cohn, F.~Yang, K.~Najafi, B.~Jones, and J.~K. Freericks, ``Minimal effective
  gibbs ansatz: A simple protocol for extracting an accurate thermal
  representation for quantum simulation,'' {\em Physical Review A}, vol.~102,
  no.~2, p.~022622, 2020.

\bibitem{sun2021quantum}
S.-N. Sun, M.~Motta, R.~N. Tazhigulov, A.~T. Tan, G.~K.-L. Chan, and A.~J.
  Minnich, ``Quantum computation of finite-temperature static and dynamical
  properties of spin systems using quantum imaginary time evolution,'' {\em PRX
  Quantum}, vol.~2, no.~1, p.~010317, 2021.

\bibitem{hao2019q}
G.~Hao~Low, N.~P. Bauman, C.~E. Granade, B.~Peng, N.~Wiebe, E.~J. Bylaska,
  D.~Wecker, S.~Krishnamoorthy, M.~Roetteler, K.~Kowalski, {\em et~al.},
  ``{Q}\# and {NWChem}: Tools for scalable quantum chemistry on quantum
  computers,'' {\em arXiv e-prints}, pp.~arXiv--1904, 2019.

\bibitem{rice2002condense}
T.~Rice, ``To condense or not to condense,'' {\em Science}, vol.~298, no.~5594,
  pp.~760--761, 2002.

\end{thebibliography}

\section*{Acknowledgements}
We would like to thank Nobuyuki Kurita, Hidekazu Tanaka, et al. for giving us access to the experimental data they obtained in Ref. \cite{kurita2019localized}. We would like to thank Travis Humble for his overall support for this project and the early discussions toward its inception. All authors, except for GH, and the research as a whole were supported by the Quantum Science Center (QSC), a National Quantum Science Initiative of the Department Of Energy (DOE), managed by Oak Ridge National Laboratory (ORNL). GH was supported by the DOE Office of Science, Basic Energy Sciences, under Contract No. DE-SC0022986. ZH acknowledges initial support from the LANL Mark Kac Fellowship and subsequent support from the Sandoz Family Foundation-Monique de Meuron program for Academic Promotion. PK additionally thanks Travis Humble for funding through the DOE Early Career Award (DOE-ECA). NME wants to thank IBM Research for their support through an internship over the summer of 2021. We acknowledge the use of IBM Quantum services for this work. This research used resources of the Oak Ridge Leadership Computing Facility, which is a DOE Office of Science User Facility supported under Contract No. DE-AC05-00OR22725. We thank Ryan Landfield (ORNL) for facilitating the process of reserving time on IBM-Q backends for the production of the results.
\section*{Author Contributions}
AB, AS, and JC conceived the project. NME performed all of the simulations and data analysis, with input from JC, ZH, and MM.  JG performed the REFF parameter optimization and ansatz diagonalization, with inputs from LC and ZH. GH, PK, and AB helped with the connection with inelastic neutron scattering data. NME. produced the first draft with input from JG, ZH, and AB. All authors contributed to the production of the final manuscript. 
\section*{Competing Interests}
Authors declare that they have no competing interests.
\section*{Data and Materials Availability}
Data is available upon request.
\clearpage
\onecolumngrid
\section{Supplementary materials}
\beginsupplement
\subsection{Vector representation of the triplet and singlet states}\label{vec}
One spin-$1/2$ state is represented as either an up state $\ket{\uparrow}$ or down state $\ket{\downarrow}$. Moreover, the singlet and triplet states of dimers originally written in their $\ket{s,m}$ form can be represented as so:
\begin{equation}\label{triplet}
  \begin{cases}
    \begin{aligned}
      \ket{1,1}\hspace{8pt} &= \ket{\uparrow \uparrow} \\
      \ket{1,0}\hspace{8pt} &= \frac{1}{\sqrt{2}}(\ket{\uparrow \downarrow} + \ket{\downarrow \uparrow)} \\
      \ket{1,-1} &= \ket{\downarrow \downarrow}
    \end{aligned}
  \end{cases} s = 1 \hspace{2pt} \text{(triplet)}
\end{equation}
\begin{equation}\label{singlet}
  \begin{cases}
    \begin{aligned}
      \ket{0,0} \hspace{8pt}= \frac{1}{\sqrt{2}}(\ket{\uparrow \downarrow} - \ket{\downarrow \uparrow})
    \end{aligned}
  \end{cases} s = 0 \hspace{2pt} \text{(singlet)}
\end{equation}
The vector representation of $\ket{\uparrow}$ is $\begin{pmatrix}
        1 \\
        0 
        \end{pmatrix}$ while that of $\ket{\downarrow}$ is $\begin{pmatrix}
        0 \\
        1 
        \end{pmatrix}$. To obtain the vector representation of the triplet and singlet states as written in Eqs.\eqref{triplet} and \eqref{singlet}, we do the following:
\begin{itemize}
    \item $\ket{\uparrow\uparrow}$, $s = 1$, $s_z = 1$\\
    \begin{equation}
        \ket{\uparrow} \otimes  \ket{\uparrow} = \begin{pmatrix}
        1 \\
        0 
        \end{pmatrix} \otimes \begin{pmatrix}
        1 \\
        0 
        \end{pmatrix} = \begin{pmatrix}
        1 \\
        0 \\
        0 \\
        0
        \end{pmatrix}
    \end{equation}
    \item $\frac{1}{\sqrt{2}}(\ket{\uparrow\downarrow} + $$\ket{\downarrow\uparrow})$, $s = 1$, $s_z = 0$ \\
    \begin{equation}
        \begin{split}
        \frac{1}{\sqrt{2}}(\ket{\uparrow} \otimes  \ket{\downarrow} + \ket{\downarrow} \otimes  \ket{\uparrow}) =  \frac{1}{\sqrt{2}}  \Bigg[ \begin{pmatrix}
        1 \\
        0 
        \end{pmatrix}  \otimes  \begin{pmatrix}
        0 \\
        1
        \end{pmatrix} +
        \begin{pmatrix}
        0 \\
        1 
        \end{pmatrix} \otimes \begin{pmatrix}
        1 \\
        0
        \end{pmatrix} \Bigg]
        = 
        \frac{1}{\sqrt{2}} \begin{pmatrix}
        0 \\
        1 \\
        1 \\
        0
        \end{pmatrix}
        \end{split}
    \end{equation}
    \item $\ket{\downarrow\downarrow}$, $s = 1$, $s_z = -1$\\
    \begin{equation}
        \ket{\downarrow} \otimes  \ket{\downarrow}= \begin{pmatrix}
        0 \\
        1 
        \end{pmatrix} \otimes \begin{pmatrix}
        0 \\
        1 
        \end{pmatrix} = \begin{pmatrix}
        0 \\
        0 \\
        0 \\
        1
        \end{pmatrix}
    \end{equation}
    \item $\frac{1}{\sqrt{2}}(\ket{\uparrow\downarrow}$ - $\ket{\downarrow\uparrow})$, $s = 0$, $s_z = 0$ \\
    \begin{equation}
        \begin{split}
            \frac{1}{\sqrt{2}}(\ket{\uparrow} \otimes  \ket{\downarrow} - \ket{\downarrow} \otimes  \ket{\uparrow}) =  \frac{1}{\sqrt{2}} \Bigg[ \begin{pmatrix}
            1 \\
            0 
            \end{pmatrix} \otimes \begin{pmatrix}
            0 \\
            1
            \end{pmatrix} -
            \begin{pmatrix}
            0 \\
            1 
            \end{pmatrix} \otimes \begin{pmatrix}
            1 \\
            0
            \end{pmatrix} \Bigg]
            = \frac{1}{\sqrt{2}} \begin{pmatrix}
            0 \\
            1 \\
            -1 \\
            0
            \end{pmatrix}
        \end{split}
    \end{equation}
\end{itemize}

\subsection{Hamiltonian models simulated}
The Hamiltonian model that we study in this paper (setting $\hbar/2 = 1$) is the 1D Heisenberg Hamiltonian model, which can be written as so:
    \begin{equation}
    H =  \sum\limits_{j=1}^{n - 1} J_{xx}\sigma^{x}_{j}\sigma^{x}_{j+1} + J_{yy}\sigma^{y}_{j}\sigma^{y}_{j+1} + J_{zz}\sigma^{z}_{j}\sigma^{z}_{j+1} +
    h \sum\limits_{j=1}^{n} \sigma^{z}_{j} \, ,
   \end{equation}
where $n$ is the number of system qubits. 

In the case of $n = 2$, we can study dimers by setting the initial state of the system qubits as one of the cases of the triplet states or singlet state as shown in Sec.~\ref{vec}.
The action of the spin-$1/2$ operators on the state $|m_1 m_2 \rangle$ is as following:\\
\begin{itemize}
    \item $\sigma^{z}_{1} |m_1 m_2 \rangle = m_1 |m_1 m_2 \rangle $
    \item $\sigma^{z}_{2} |m_1 m_2 \rangle = m_2 |m_1 m_2 \rangle $
    \item $\sigma^{x}_{1} |m_1 m_2 \rangle = (\sigma^{+}_{1} + \sigma^{-}_{1}) |m_1 m_2 \rangle$
    \item $\sigma^{x}_{2} |m_1 m_2 \rangle = (\sigma^{+}_{2} + \sigma^{-}_{2}) |m_1 m_2 \rangle$
    \item $\sigma^{y}_{1} |m_1 m_2 \rangle = \frac{1}{i}(\sigma^{+}_{1} - \sigma^{-}_{1}) |m_1 m_2 \rangle$
    \item $\sigma^{y}_{2} |m_1 m_2 \rangle = \frac{1}{i}(\sigma^{+}_{2} - \sigma^{-}_{2}) |m_1 m_2 \rangle$
\end{itemize}
where the complete action of the operators $\sigma^{x}_{1} \sigma^{x}_{2}$ and $\sigma^{y}_{1} \sigma^{y}_{2}$ found in our model for $n = 2$ is as follows:
\begin{equation}
\centering
\begin{split}
\sigma^{x}_{1} \sigma^{x}_{2}|m_1 m_2 \rangle  = (\sigma^{+}_{1} + \sigma^{-}_{1})(\sigma^{+}_{2} + \sigma^{-}_{2})  |m_1 m_2 \rangle = \\
\Bigg[\sqrt{3/4 -m_2(m_2 +1)}\sqrt{3/4 - m_1(m_1 + 1)} |m_1+1 \hspace{2pt} m_2+1 \rangle \\
+ \sqrt{3/4 -m_2(m_2 -1)}\sqrt{3/4 - m_1(m_1 + 1)} |m_1+1 \hspace{2pt} m_2-1 \rangle \\
+ \sqrt{3/4 -m_2(m_2 +1)}\sqrt{3/4 - m_1(m_1 - 1)} |m_1-1 \hspace{2pt} m_2+1 \rangle \\
+ \sqrt{3/4 -m_2(m_2 -1)}\sqrt{3/4 - m_1(m_1 - 1)} |m_1-1 \hspace{2pt} m_2-1 \rangle \Bigg] \, ,
\end{split}
\end{equation}
\begin{equation}
\centering
\begin{split}
\sigma^{y}_{1} \sigma^{y}_{2}|m_1 m_2 \rangle  = -(\sigma^{+}_{1} - \sigma^{-}_{1})(\sigma^{+}_{2} - \sigma^{-}_{2})  |m_1 m_2 \rangle = \\
- \Bigg[\sqrt{3/4 -m_2(m_2 +1)}\sqrt{3/4 - m_1(m_1 + 1)} |m_1+1 \hspace{2pt} m_2+1 \rangle \\
- \sqrt{3/4 -m_2(m_2 -1)}\sqrt{3/4 - m_1(m_1 + 1)} |m_1+1 \hspace{2pt} m_2-1 \rangle \\
- \sqrt{3/4 -m_2(m_2 +1)}\sqrt{3/4 - m_1(m_1 - 1)} |m_1-1 \hspace{2pt} m_2+1 \rangle \\
+ \sqrt{3/4 -m_2(m_2 -1)}\sqrt{3/4 - m_1(m_1 - 1)} |m_1-1 \hspace{2pt} m_2-1 \rangle \Bigg] \, ,
\end{split}
\end{equation}
From this, we can find the matrix representation of the Hamiltonian model as follows:
\begin{equation}
    \langle \hat{H} \rangle  = \begin{pmatrix}
2h + J_{zz} & 0 & 0 & J_{xx} - J_{yy}\\
0 & -J_{zz} & J_{xx} + J_{yy} & 0 \\
0 & J_{xx} + J_{yy} & -J_{zz} & 0 \\
J_{xx} - J_{yy} & 0 & 0 & -2h + J_{zz} 
\end{pmatrix} .
\end{equation}
\subsection{Lehmann representation of the correlation functions}\label{Lehmann}
We obtain the Lehmann representation of the correlation functions in the following manner:
\begin{equation}
    \begin{split}
        \langle \sigma^{\alpha}_{i}(t) \sigma^{\beta}_{j} \rangle = \langle\psi|\sigma^{\alpha}_{i}(t)\sigma^{\beta}_{j}(0)|\psi\rangle =& \\\langle\psi|e^{i H t}\sigma^{\alpha}_{i}e^{- i H t}|p\rangle\langle p|\sigma^{\beta}_{j}(0)|\psi\rangle=& \\\sum_{p} e^{-iE_{p}t}\langle\psi|e^{i H t}\sigma^{\alpha}_{i}|p\rangle\langle p|\sigma^{\beta}_{j}(0)|\psi\rangle=&
        \\ \sum_{p,r} e^{-iE_{p}t}e^{iE_{r}t}\langle\psi|r\rangle \langle r| \sigma^{\alpha}_{i}|p\rangle\langle p|\sigma^{\beta}_{j}(0)|\psi\rangle
    \end{split}
\end{equation}
For the local (same-site) correlation functions with $\alpha = \beta$, we can write the Lehmann representation as follows:
\begin{equation}
    \begin{split}
        C^{\alpha \alpha}_{i,i}(t)= \langle\psi_0|e^{iHt}\sigma^{\alpha}_{i}e^{-iHt}\sigma^{\alpha}_{i}|\psi\rangle =&  \\ e^{iE_{0}t}\langle\psi_0|\sigma^{\alpha}_{i}e^{-i H t}\sigma^{\alpha}_{i}|\psi\rangle=&
        \\ \sum_{n} e^{iE_{0}t}e^{-iE_{n}t}|\langle\psi_0|\sigma^{\alpha}_{i}|\psi\rangle|^2
    \end{split}
\end{equation}
As shown in Eqs.\eqref{triplet} and \eqref{singlet}, we can represent the triplet and singlet states for a 2 spin-1/2 system. These states are the eigenstates of the 1D Heisenberg Hamiltonian model for two spins, with the following eigenvalues (assuming that $J_{xx} = J_{yy} = J_{zz} = J$):
\begin{table}[h!]
\centering
\begin{tabular}{||c | c||} 
 \hline
State & Energy  \\ [0.5ex] 
 \hline\hline
$\ket{\psi_0}=\frac{1}{\sqrt{2}}(\uparrow \downarrow - \downarrow \uparrow)$ & $E_0 = -3J$  \\ 
 $\ket{\psi_1} = \ \uparrow \uparrow $ & $E_1 = J - 2h$  \\
 $\ket{\psi_2} = \frac{1}{\sqrt{2}}(\uparrow \downarrow + \downarrow \uparrow)$ & $E_2 = J$ \\
 $\ket{\psi_3} = \ \downarrow \downarrow$ & $E_3 = J + 2h$ \\ [1ex] 
 \hline
\end{tabular}
\label{table:state}
\end{table}

For $\alpha = x,y,z$, we get the following cases:
\begin{itemize}
    \item $\alpha = x$
    \begin{equation}
        |\langle\psi_0 |\sigma^x _i|\psi_n \rangle|^2 = \frac{1}{2}, \hspace{3pt} \text{for} \hspace{3pt}  n =1 \hspace{3pt} \text{and} \hspace{3pt} 3 \, ,
    \end{equation}
hence,
    \begin{equation} \label{cxx11}
        C^{xx}_{i,i}(t) = \frac{1}{2}[e^{-i(4J + 2h)t} + e^{-i(4J - 2h)t}] \, ,
    \end{equation}
        \item $\alpha = y$
    \begin{equation}
        |\langle\psi_0 |\sigma^y _i|\psi_n \rangle|^2 = \frac{1}{2}, \hspace{3pt} \text{for} \hspace{3pt}  n =1 \hspace{3pt} \text{and} \hspace{3pt} 3 \, ,
    \end{equation}
hence,
    \begin{equation} \label{cyy11}
        C^{yy}_{i,i}(t) = \frac{1}{2}[e^{-i(4J + 2h)t} + e^{-i(4J - 2h)t}] \, ,
    \end{equation}
            \item $\alpha = z$
    \begin{equation}
        |\langle\psi_0 |\sigma^z _i|\psi_n \rangle|^2 = 1, \hspace{3pt} \text{for} \hspace{3pt}  n =0 \hspace{3pt} \text{and} \hspace{3pt} 2 \, ,
    \end{equation}
hence,
    \begin{equation}\label{czz11}
        C^{zz}_{i,i}(t) = e^{-i4Jt} .
    \end{equation}
\end{itemize}

\begin{figure*}[htp!]
    \centering
    \includegraphics[width=0.68\textwidth]{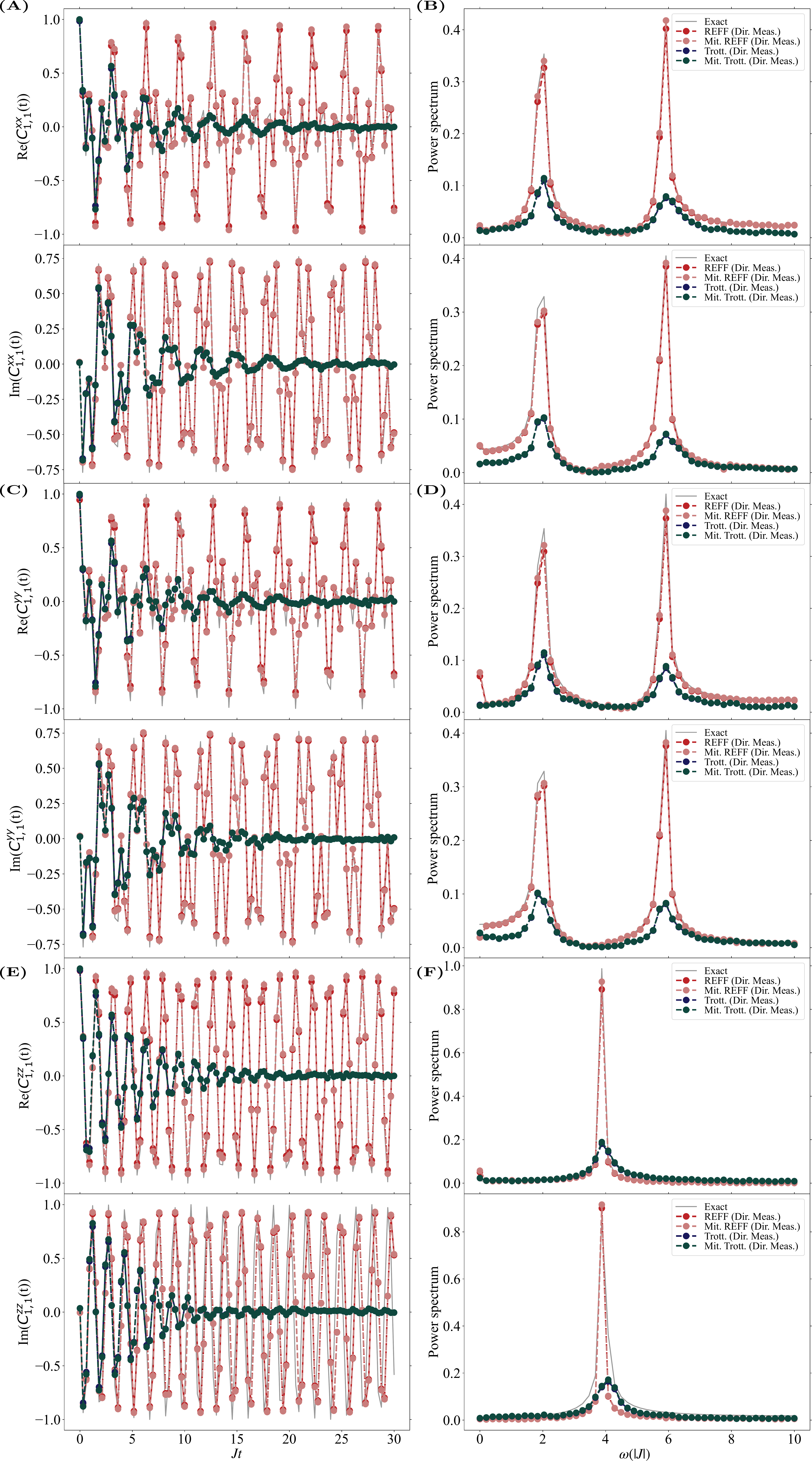}
    \caption{\textbf{Spectrum of different correlation functions of the 2 spin-1/2 Heisenberg model system}: We set $J = 1$ and $h = 1$ for these results. Both Trotterization and REFF were used to simulate the time evolution and direct measurements were implemented. The use of measurement error mitigation is also presented. Little difference is made when applying it, however, in this case. \textbf{(A)},\textbf{(C)} and \textbf{(E)} are the plots of the real and imaginary parts of the correlation functions $C^{xx}_{1,1}(t)$, $C^{yy}_{1,1}(t)$, and $C^{zz}_{1,1}(t)$ against $(J = 1)t$ respectively, while \textbf{(B)},\textbf{(D)} and \textbf{(F)} are the plots of their corresponding Fast Fourier transforms (representing their power spectra) against frequency $\omega$. We see that we are able to obtain the expected peaks in the power spectrum plots that were calculated in Eqs.\eqref{cxx11},\eqref{cyy11}, and \eqref{czz11}. The results were obtained using \textit{ibm\_auckland}, with the qubits' initial layout of [4,1] for the direct measurements. $t = 30$, the number of time steps $= 100$, $dt = 0.3$, and the number of shots $= 8000$.}.
    \label{spec_pic}
\end{figure*}

When observing the spectrum of these individual correlation functions, we get peaks at the corresponding energies of each time evolution operator included in the calculations of $C^{xx}_{i,i}(t)$, $C^{yy} _{i,i}(t)$, and $C^{zz}_{i,i}(t)$. We show this in Fig.\ref{spec_pic}.

\subsection{Generalization of the CNOT gate: C-$U$ gate}
In the process of measuring the correlation functions, we make use of the controlled-$U$ (C-$U$) gate, where U is a unitary operator acting on a multi-qubit state $|\Psi_s\rangle$. This generalization of the CNOT gate can be represented in the following form \cite{somma2002simulating}:
\begin{equation}\label{CNOT}
\text{C}-U
  \begin{cases}
    \begin{aligned}
     \ket{0}_\text{a} \otimes \ket{\Psi_s} &\rightarrow \ket{0}_\text{a} \otimes \ket{\Psi_s} 
      \\
      \ket{1}_\text{a} \otimes \ket{\Psi_s} &\rightarrow \ket{1}_\text{a} \otimes [U\ket{\Psi_s}]
    \end{aligned}
  \end{cases} 
\end{equation}
For $U(t) = e^{-i \hat{Q} t}$, where $\hat{Q}$ is Hermitian, the operational representation of the C-$U$ gate is $U(t/2)U(t/2)^{-\sigma^{\text{a}}_z}$ ($U(t)^{\sigma^{\text{a}}_{z}} = e^{i\hat{Q}\otimes \sigma^{\text{a}}_{z} t}$), where a is the control (or in this case, ancillary) qubit.
\subsection{Methods}
\subsubsection{Trotterization}
According to the \textbf{Baker-Campbell-Hausdorff formula}, the product of two exponentials is written in the following manner:
\begin{equation}
    e^X e^Y = e^Z \, ,
\end{equation}
where $Z$ is written as:
\begin{equation}
    Z = X + Y + \frac{1}{2}[X,Y] + \frac{1}{12}[X,[X,Y]] - \frac{1}{12}[Y,[X,Y]] + \dots \, ,
\end{equation}
If $X$ and $Y$ are two commuting operators, then the expression reduces to $e^X e^Y = e^{X+Y}$. However, this is not the case for non-commuting terms as is evident. Hence, for a Hamiltonian operator $\hat{H}$ written as the sum of non-commuting operators, it is not an easy task to factor out the time evolution operator into unitary operators that can be translated into quantum gates. However, we can implement what is known as the \textbf{Suzuki-Trotter decomposition} to transform the time evolution operator into a form implementable on a quantum device. This is an approximation method that incurs what is known as a Trotter error that can be changed based on the order of the approximation in which individual term unitaries are applied and the number of repetitions of applying the sequence of gates representing our one Trotter step, i.e. the Trotter number \cite{hao2019q}.
However, such changes can increase the circuit depth, making the implementation of such a process impractical.\\
The Trotterization approximation comes from the fact that a Hamiltonian $H$ that is written as a sum of operators (i.e., $H_1 + H_2$) can be described by the Lie product formula:
\begin{equation}
    e^{i(H_1 + H_2)t} = \lim_{N\rightarrow \infty} (e^{-iH_1 t/N}e^{-iH_2 t/N})^N \, ,
\end{equation}
Given that the limit of this formula is infinite, we must truncate the series when implementing this formula on a quantum device. The truncation introduces error in the simulation that we can bound by a maximum simulation error $\epsilon$ such that $||e^{-iHt} - U|| \leq \epsilon$. This truncation is the Trotterization approximation and based on when the truncation is taken, the order of the Trotterization is determined.
\subsubsection{Resource Efficient Fast-forwarding (REFF)}
\begin{figure}[ht!]
    \centering
    \includegraphics[width=0.75\textwidth]{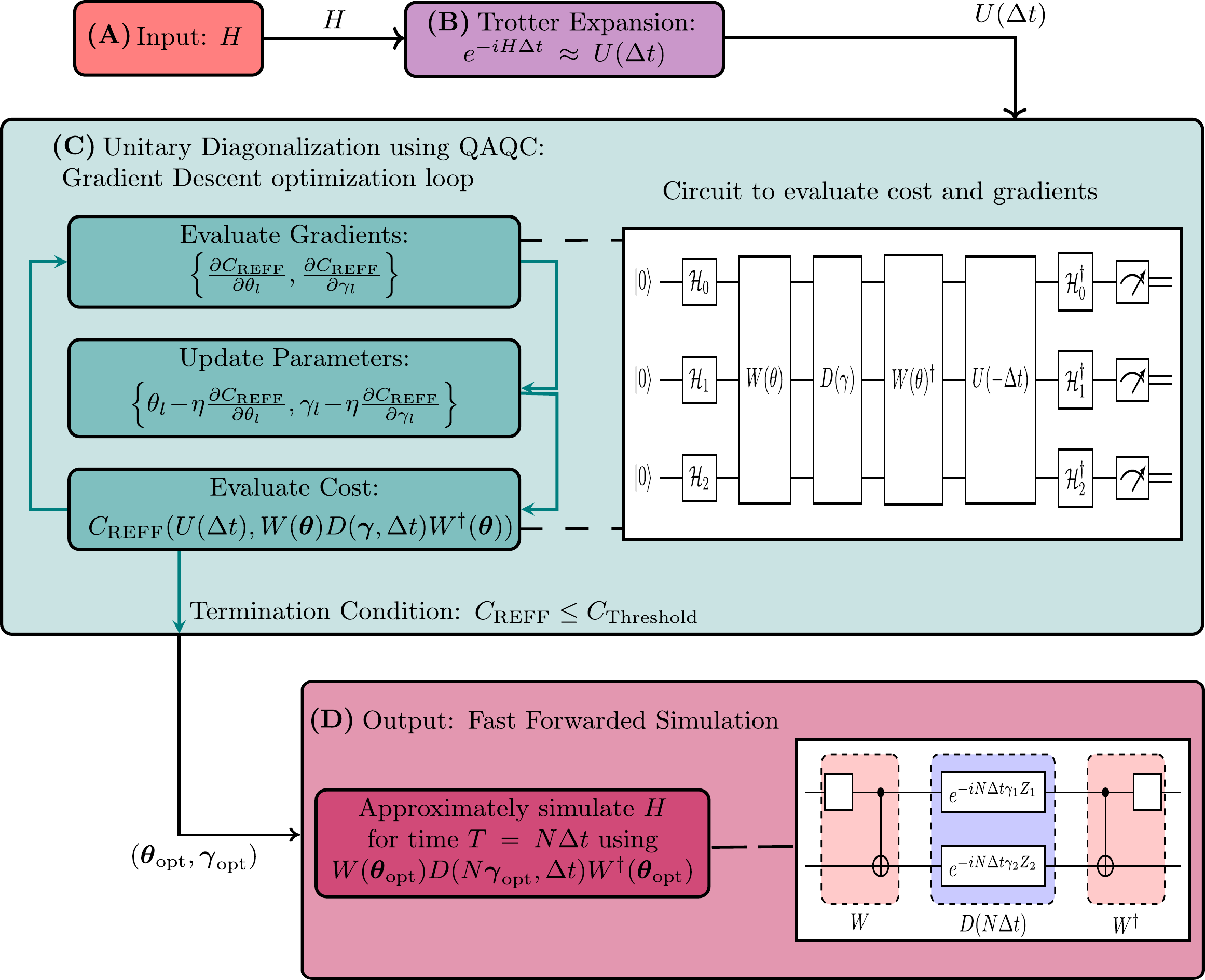}
    \caption{\textbf{The REFF algorithm}: \textbf{(A)} From the input Hamiltonian $H$, \textbf{(B)} a gate sequence representing a single-timestep Trotterized unitary $U(\Delta t)$ is obtained. \textbf{(C)} Such unitary is subsequently variationally diagonalized by fitting a parameterized factorization in the form of $V_{\Delta t}(\boldsymbol{\alpha}) = W(\boldsymbol{\theta}) D_{\Delta t}(\boldsymbol{\gamma}) W^{\dagger}(\boldsymbol{\theta})$. This variational subroutine employs gradient descent to minimize the cost function $C_{\text{REFF}}$ through the means of efficiently estimating its gradient with the shown short-depth circuit. $\mathcal{H}_k$ is a Haar-random single qubit unitary. The optimization loop continues until the termination condition in Eq.\eqref{thresh} is met. \textbf{(D)} Once the termination condition is met, the optimal parameters ($\boldsymbol{\theta}_{\text{opt}}, \boldsymbol{\gamma}_{\text{opt}}$) are used to implement the fast-forwarded time evolution simulation. The fast-forwarding error grows sublinearly with the simulation time. The fast-forwarding is performed by properly scaling the parameters of the diagonal unitary, $D_{\Delta t}(\boldsymbol{\gamma}_\text{opt}) \rightarrow D_{N\Delta t}(\boldsymbol{\gamma}_\text{opt})$.}
    \label{flowchart}
\end{figure}

As shown in Fig.\ref{flowchart}, the gradients of $C_{\text{REFF}}$ are calculated in the process of minimizing the local cost function and finding the optimized parameters. The partial derivative of $C_{\text{REFF}}(U,V(\mathbf{\theta}, \mathbf{\gamma}))$ with respect to $\theta_l$ is 
\begin{equation}
\begin{split}
    \frac{\partial C_{\text{REFF}}(U,V)}{\partial \theta_l} = \frac{1}{2}\bigg(C_{\text{REFF}}(U,W_{l+}DW^\dagger) - C_{\text{REFF}}(U,W_{l-}DW^\dagger) &\\
    + C_{\text{REFF}}(U,WD(W_{l+})^\dagger) - C_{\text{REFF}}(U,WD(W_{l-})^\dagger)\bigg) \, ,
\end{split}
\end{equation}\\
where the unitary $W_{l+} (W_{l-})$ is generated from the original unitary $W(\theta)$ by the addition of an extra $\frac{\pi}{2} (\frac{-\pi}{2})$ rotation about a given parameter's rotation axis:
\begin{equation}
    W_{l \pm} := W(\thv_{l\pm})_i \hspace{7pt} \text{with}\hspace{7pt} (\thv_{l\pm})_i := \theta_i \pm \frac{\pi}{2}\delta_{i,l}\, ,
\end{equation}
The analogous formula for the partial derivative with respect to $\gamma_l$ is
\begin{equation}
    \frac{\partial C_{\text{REFF}}(U,V)}{\partial \gamma_l} = \frac{1}{2}\bigg(C_{\text{REFF}}(U,WD_{l+}W^\dagger) - C_{\text{REFF}}(U,WD_{l-}W^\dagger)\bigg)
\end{equation}

This variational loop is exited when the termination condition $C_{\text{REFF}} \leq C_{\text{Threshold}}$ is reached, with
    \begin{equation}\label{thresh}
        C_\text{Threshold} \approx \frac{\epsilon}{16n_{\rm targ}^2} - \frac{\epsilon^2}{4(2^n + 1)} \, ,
    \end{equation}
    where $1-\epsilon$ is the target simulation fidelity after $n_{\rm targ}$ fast-forwarding steps.
Upon reaching the termination condition, the optimal parameters $\boldsymbol{\alpha}_\text{opt} = (\boldsymbol{\theta}_\text{opt}, \boldsymbol{\gamma}_\text{opt})$ are used to implement the fast-forwarded simulation $W(\thv)D_{N\Delta t}(\gamv)W^\dagger(\thv)$, with the fast-forwarding error growing quadratically in the simulation time. 

\begin{figure}[ht!]
    \centering
    \includegraphics[width=1\textwidth]{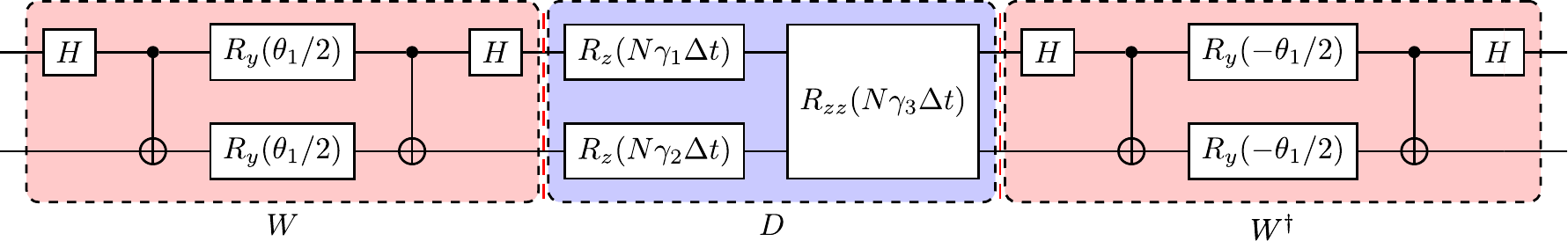}
    \caption{The REFF ansatz for the 2-site Heisenberg model. $\theta_1$, $\gamma_1$, $\gamma_2$ and $\gamma_3$ are the optimized parameters. The same ansatz is used for the XY + ZZ perturbation model simulation, with different optimized parameters.}
    \label{ansatz}
\end{figure}

An example of an REFF ansatz used for the 2-site Heisenberg model is shown in Fig.\ref{ansatz}. The optimization process of the parameters for this ansatz was classically performed. Optimization on hardware can be implemented, however.
\subsubsection{Direct measurements}\label{DMeas}
In attempt to estimate $\langle\psi |W^{\dagger} U W G| \psi \rangle$ as shown in Fig.\ref{fig:DirectIndirectMeth}, the following quantities are used:
\begin{equation}\label{upm}
    \langle U \rangle _{\pm} =\langle \psi| e^{\mp i \pi G/4} W^\dagger U W e^{\pm i \pi G/4}|\psi \rangle \, ,
\end{equation}
\begin{equation}\label{umg}
    \langle U \rangle _{M_{G} = \pm 1} = \frac{1}{4p(M_G = \pm 1)} \langle \psi | (I \pm G) W^\dagger U W  (I \pm G) |\psi \rangle \, ,
\end{equation}
where $p(M_G = \pm 1)$ is the probability of getting the measurement $M_G = \pm 1$ by performing $\mathcal{M}_G$ on $\ket{\psi}$; $p(M_G = \pm 1) = \abs{\abs{\frac{1}{2} (I \pm G) \ket{\psi}}}^2$. Using the expressions in Eq.\eqref{upm} and Eq.\eqref{umg}, we can estimate  $\langle\psi |W^{\dagger} U W G| \psi \rangle$ as:
\begin{equation}
    \begin{split}
     \langle\psi |W^{\dagger} U W G| \psi \rangle = 
      p(M_G = +1)\langle U \rangle_{M_G = +1}  - p(M_G = -1)\langle U \rangle_{M_G = -1}- \frac{i}{2}(\langle U \rangle_+ - \langle U \rangle_-) \, ,
     \end{split}
\end{equation}
where $p(M_G = +1)\langle U \rangle_{M_G = +1}  - p(M_G = -1)\langle U \rangle_{M_G = -1}$ represents the real part of the coefficient and $-\frac{1}{2}(\langle U \rangle_+ - \langle U \rangle_-)$ represents the imaginary part of the coefficient respectively up to a phase. 
\begin{figure}[ht]
    \centering
    \includegraphics[width=1\linewidth]{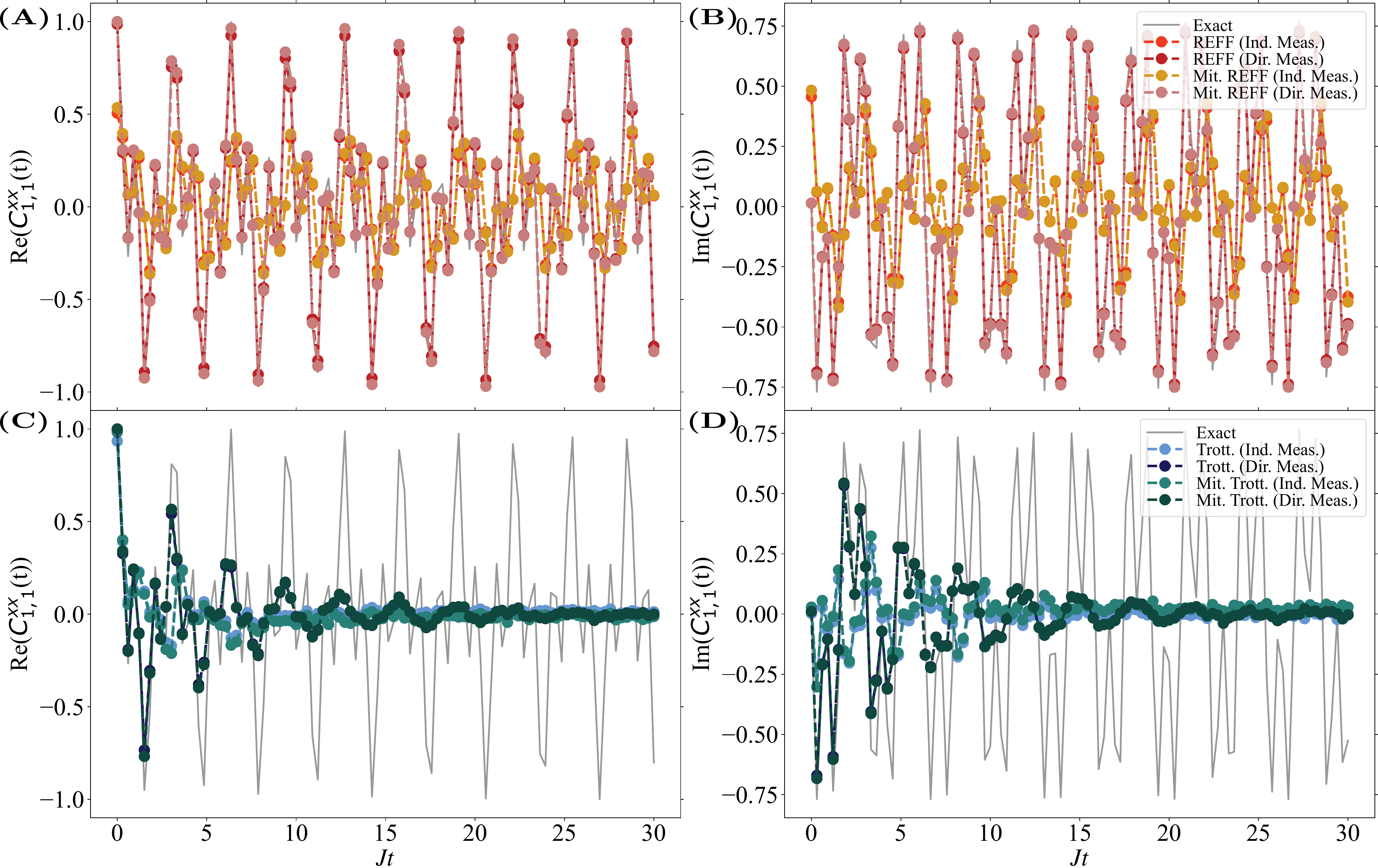}
\caption{\textbf{Application of different methods to accurately measure the correlation functions:} In the different panels, results of measuring the real and imaginary parts of the correlation function $C^{xx}_{1,1}(t)$ over time are presented respectively. The Hamiltonian employed is the Heisenberg model for a 2 spin-1/2 system.  \textbf{(A)},\textbf{(B)} Show the Trotterization results while \textbf{(C)},\textbf{(D)} Show the REFF results. We can see how the trend of the results stays mostly constant over time. We can see the effect of making use of all of the methods we discussed in this paper from time evolution simulation techniques (Trotterization vs. REFF), measurement methods (indirect vs. direct), and applying measurement error mitigation. The starkest difference comes when applying REFF instead of Trotterization. Following that is the implementation of direct measurements instead of indirect measurements. Lastly comes the application of measurement error mitigation, where not much change is introduced. Results were obtained using \textit{ibm\_geneva}, with the qubits' initial layout of [1,2] for the direct measurements and [0,1,2] for the indirect measurements. $t = 30$, the number of time steps $= 100$, $dt = 0.3$, and the number of shots $= 8000$.}
    \label{all_res}
\end{figure}
\subsection{RMS errors in results}
\begin{figure*}[h!]
\centering
 \includegraphics[width=1.0\textwidth]{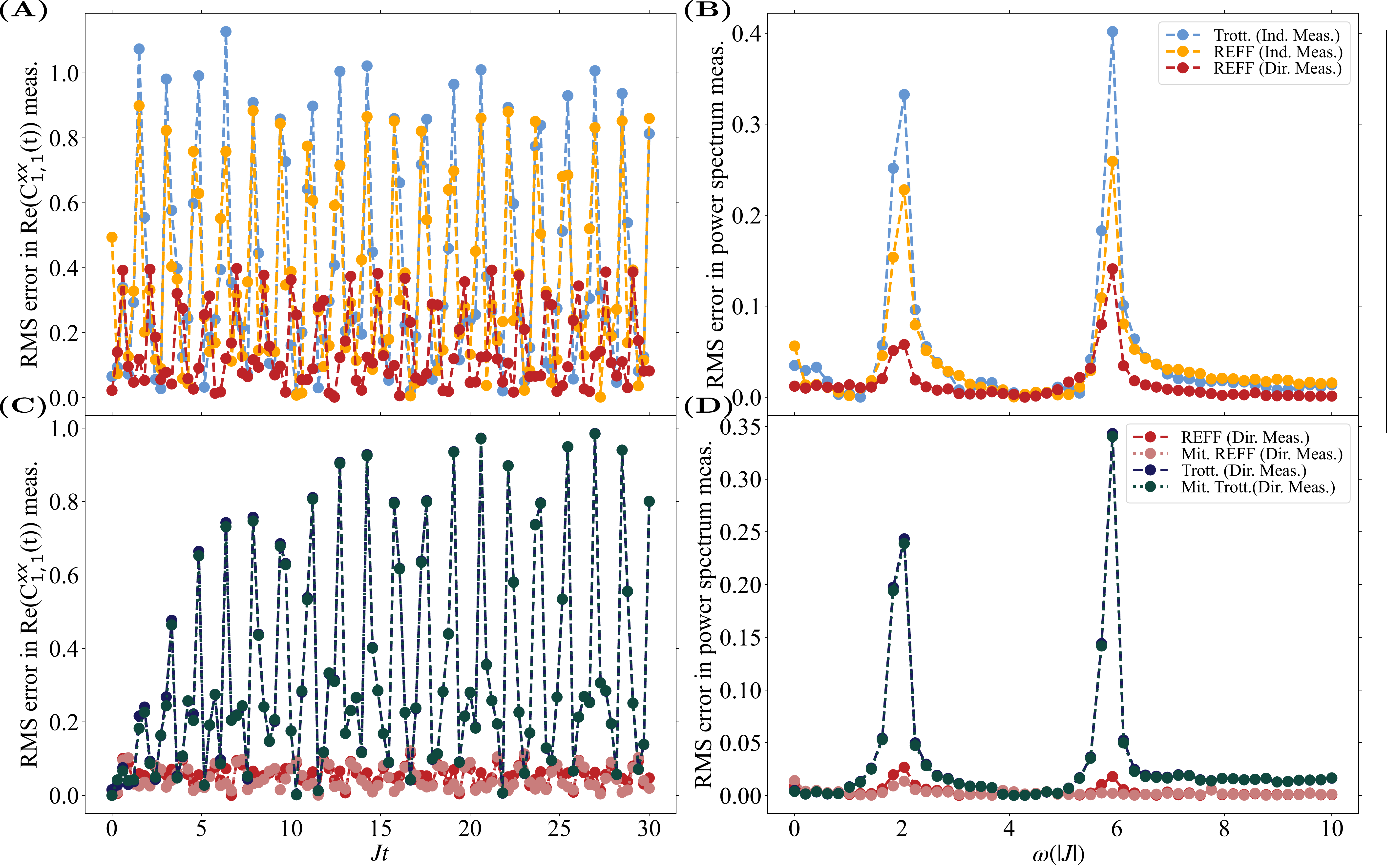}
\caption{\textbf{Error analysis of the results in Figs.\ref{mit_ex} and \ref{fig5}\textbf{(A)}-\textbf{(B)} :} \textbf{(A)} and \textbf{(B)} show the Root Mean Square (RMS) errors of the results in both the time and frequency domain for Fig.\ref{mit_ex} while \textbf{(C)} and \textbf{(D)} show the RMS errors of the results in both the time and frequency domain for Fig.\ref{fig5}\textbf{(A)}-\textbf{(B)}. We see how over time, the error rapidly increases for the Trotterization results while it stays steady for the REFF results. Measurement error mitigation slightly enhances the accuracy of the results. }
\label{err_anal1}
\end{figure*}

We present in Fig.\ref{err_anal1} the results of calculating the RMS errors of the correlation functions in Figs.\ref{mit_ex} and \ref{fig5}\textbf{(A)}-\textbf{(B)} respectively over time. Moreover, we also find the RMS error in the frequency domain for their corresponding Fourier transforms.
\subsection{IBM devices used for results}
We now present some of the calibration details of the IBM devices that were used to produce our results.
\begin{figure}[ht]
    \centering
    \includegraphics[width=0.7\linewidth]{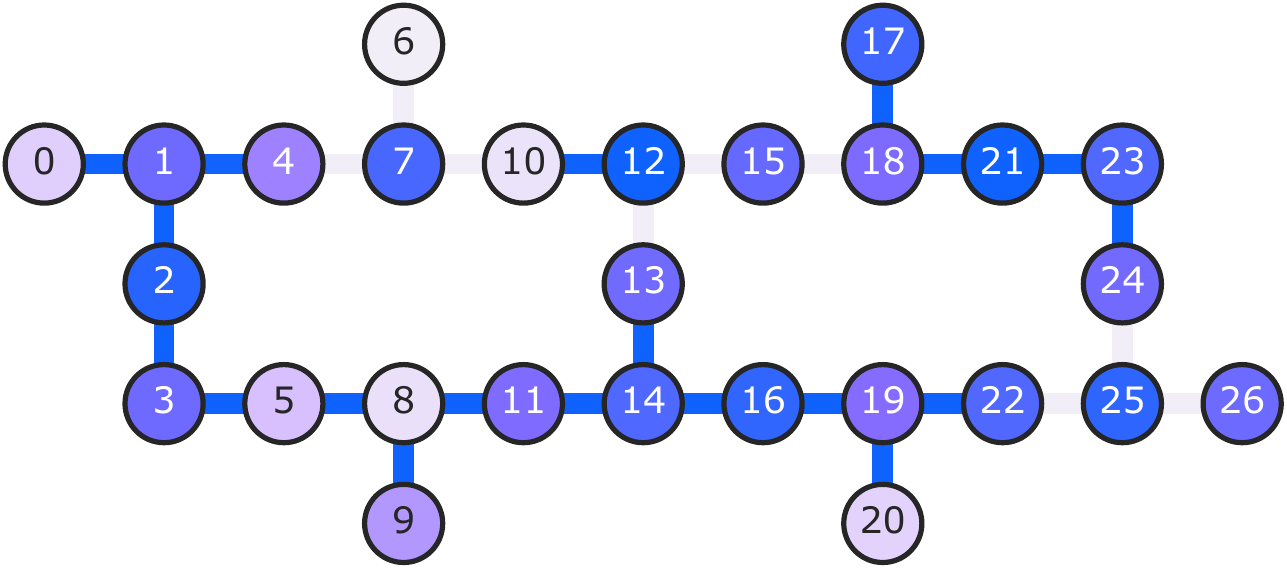}
    \caption{\textit{ibm\_geneva}: A 27-qubit device. }
    \label{geneva}
\end{figure}
\begin{table}[h!]
\centering
\begin{tabular}{||c | c | c | c||} 
 \hline
 Qubit $\#$: & Qubit 0 (ancilla) & Qubit 1 & Qubit 2 \\ [0.5ex] 
 \hline\hline
 Frequency (GHz) & 4.826 & 4.673 & 4.581\\ 
 $T_1 (\mu s)$ & 423.36 & 422.04 & 434.21 \\
 $T_2 (\mu s)$ & 447.40 & 181.55 & 294.19 \\
 Readout assignment error & 0.0157& 0.0075 & 0.0133 \\ [1ex] 
 \hline
\end{tabular}
\caption{\textbf{Calibration details of \textit{ibm\_geneva}}: Listed in this table are some of the calibration details of \textit{ibm\_geneva} when using it to produce the correlation functions for the Heisenberg model Hamiltonian for isolated dimers in Fig.\ref{mit_ex} and Fig.\ref{all_res}. Qubits 1 and 2 were the system qubits while qubit 0 was used as an ancillary qubit in the case of indirect measurements. It is also to be noted that the CNOT error between the qubits was as follows: 0-1: 0.0049; 1-2: 0.0073.}
\label{table:2}
\end{table}

\begin{table}[h!]
\centering
\begin{tabular}{||c | c | c | c||} 
 \hline
 Qubit $\#$: & Qubit 7 (ancilla) & Qubit 4 & Qubit 1 \\ [0.5ex] 
 \hline\hline
 Frequency (GHz) & 4.828 & 4.920 & 5.074\\ 
 $T_1 (\mu s)$ & 175.53 & 65.558 & 222.39 \\
 $T_2 (\mu s)$ & 218.72 & 190.53 & 190.11\\
 Readout assignment error & 0.007 & 0.0072 & 0.0196 \\ [1ex] 
 \hline
\end{tabular}
\caption{\textbf{Calibration details of \textit{ibm\_auckland}}: Listed in this table are some of the calibration details of \textit{ibm\_auckland} when using it to produce the correlation functions for the Heisenberg model Hamiltonian for isolated dimers in Fig.\ref{Heis_Int_Dir} and Fig.\ref{spec_pic}. Qubits 4 and 1 were the system qubits while qubit 7 was used as the ancillary qubit in the case of indirect measurements. It is also to be noted that the CNOT errors between the qubits were as follows: 7-4: 0.00927; 4-1: 0.00679.
}
\label{table:1}
\end{table}
\begin{figure}[h!]
    \centering
    \includegraphics[width=0.7\linewidth]{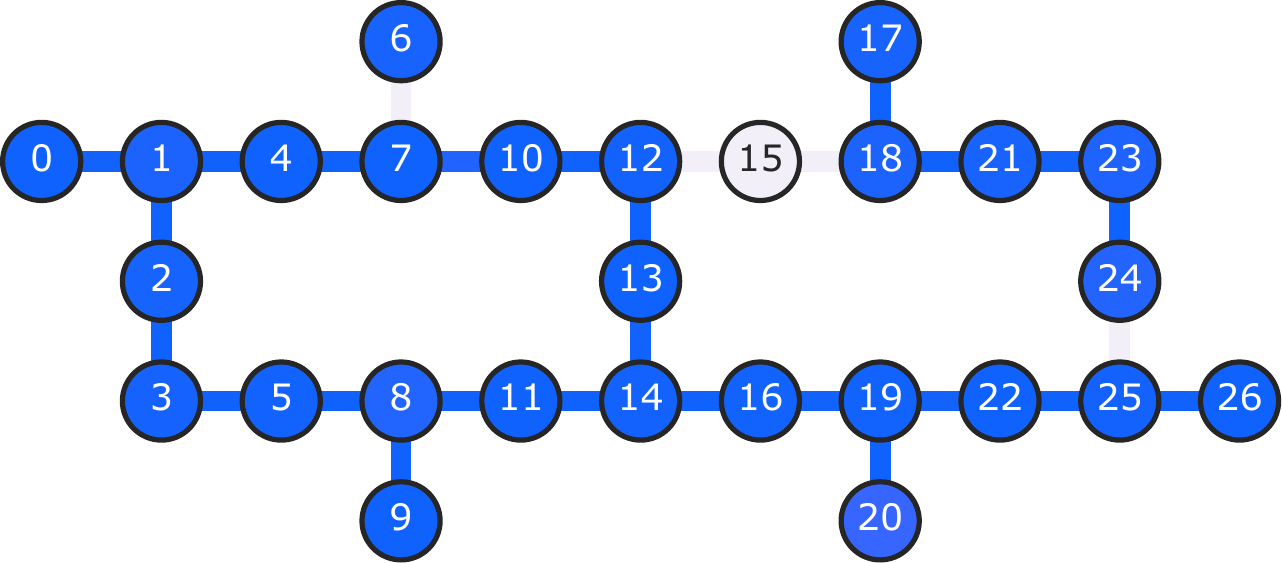}
\caption{\textit{ibm\_auckland}: A 27-qubit device.}
    \label{auckland}
\end{figure}
\begin{figure}[h!]
    \centering
    \includegraphics[width=0.7\linewidth]{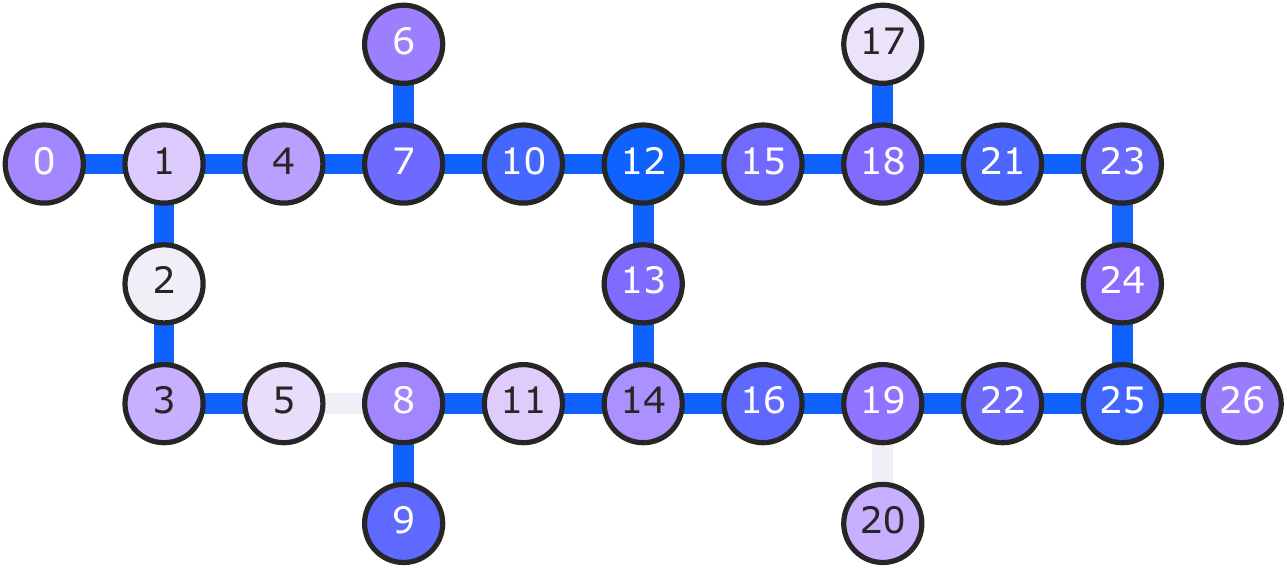}
    \caption{\textit{ibm\_hanoi}: A 27-qubit device. }
    \label{hanoi}
\end{figure}
\begin{table}[h!]
\centering
\begin{tabular}{||c | c | c||} 
 \hline
 Qubit $\#$: & Qubit 9 & Qubit 8  \\ [0.5ex] 
 \hline\hline
 Frequency (GHz) & 4.875 & 5.031 \\ 
 $T_1 (\mu s)$ & 241.44 & 178.23  \\
 $T_2 (\mu s)$ & 364.13 & 342.41 \\
 Readout assignment error & 0.0067 & 0.0113 \\ [1ex] 
 \hline
\end{tabular}
\caption{\textbf{Calibration details of \textit{ibm\_hanoi} of experimental data simulation}: Listed in this table are some of the calibration details of \textit{ibm\_hanoi} when using it to produce the correlation functions for the XY + ZZ perturbation model Hamiltonian for isolated dimers in Fig.\ref{sim_Res}. Qubits 9 and 8 were the system qubits. It is also to be noted that the CNOT error between the qubits was as follows: 9-8: 0.00658.
}
\label{table:3}
\end{table}
\subsection{The Ba$_2$CoSi$_2$O$_6$Cl$_2$ molecule}\label{compound}
On a dimer lattice, magnons are created when a magnetic field exceeding the critical field is applied \cite{rice2002condense}. Such magnons can hop to neighboring dimer sites and interact with each other via the transverse and longitudinal components of the exchange interactions, respectively. For the simplified two-dimensional case, the hopping and repulsive terms are proportional to $(J_{11}+J_{22})-(J_{12} + J_{21})$ and $J_{11} + J_{22} + J_{12} + J_{21}$ respectively. In the case of having perfect frustration of the interdimer exchange interactions, i.e., $J_{11} + J_{22} = J_{12} + J_{21}$, the hopping of magnons is completely suppressed, forming a periodic array of half-filled magnons due to the competition between the repulsive interactions and the Zeeman energy. It was found in Ref.~\cite{kurita2019localized} that the interdimer interactions in Ba$_2$CoSi$_2$O$_6$Cl$_2$ almost perfectly satisfy such a frustration condition, and hence we can treat it as a system of isolated dimers. 
\renewcommand\thesubfigure{\textbf{\Alph{subfigure}}}
 \begin{figure}[!ht]
    \subfloat[\label{Ba2Co}]
    {
    \includegraphics[width=0.6\linewidth]{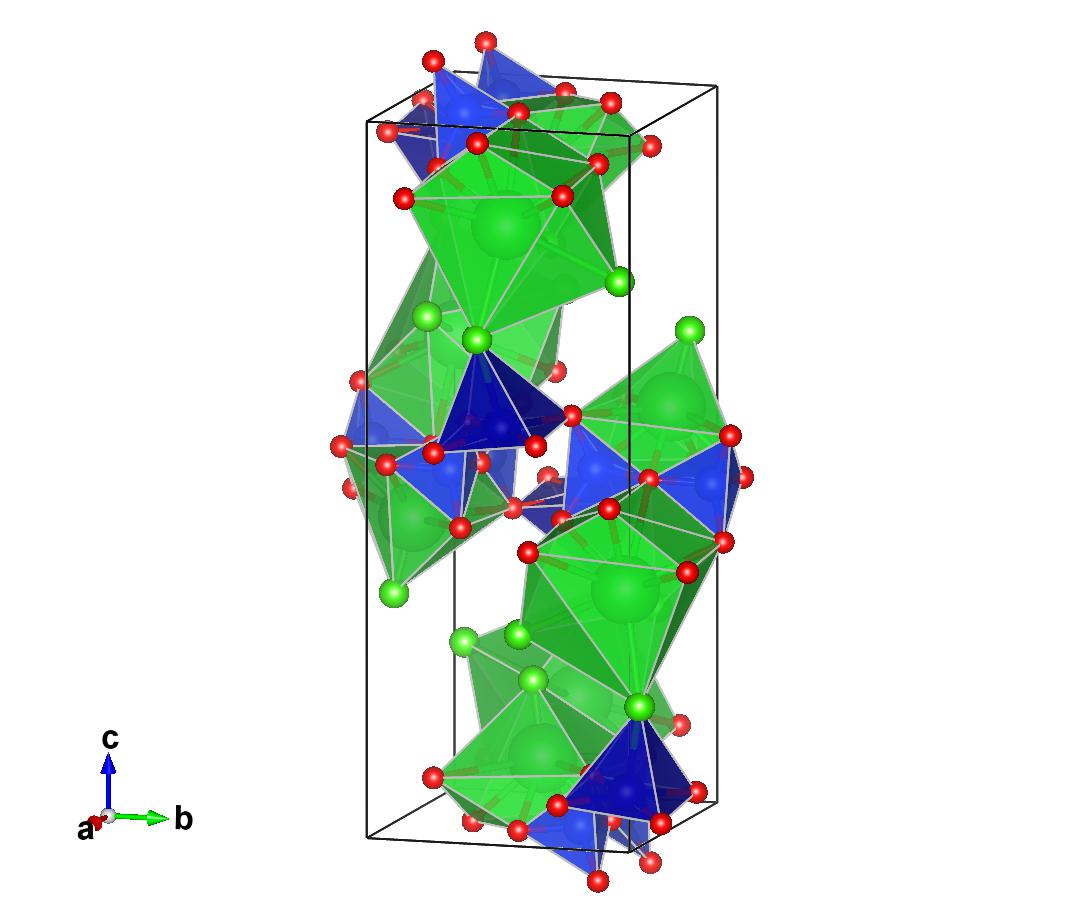}}
    \hfill
    \subfloat[\label{mag}]
    {
   \includegraphics[width=0.35\linewidth,height=9cm]{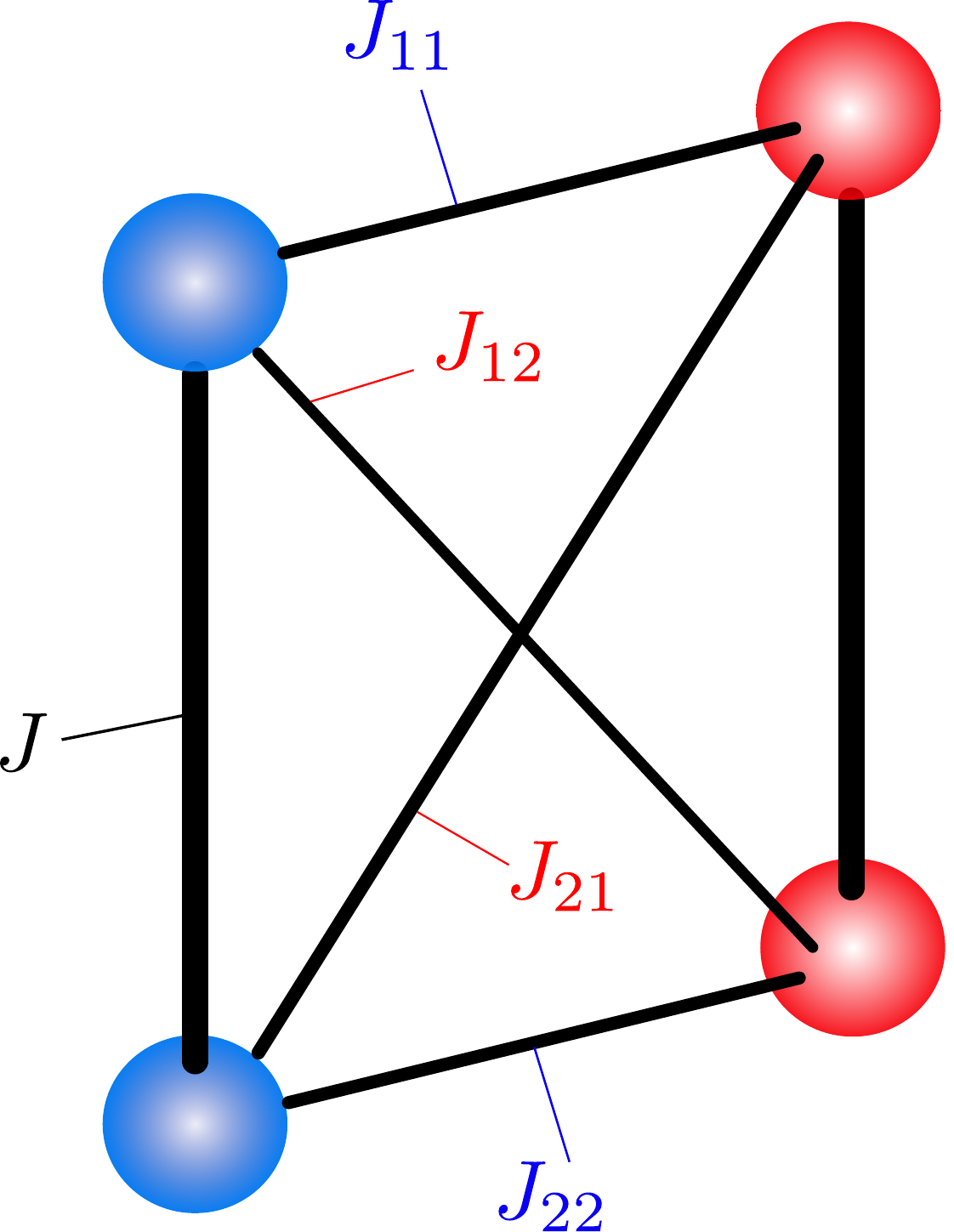} }
    \caption{\textbf{(A)} The Ba$_2$CoSi$_2$O$_6$Cl$_2$ molecule. \textbf{(B)} A schematic view of the 2D exchange network of the Ba$_2$CoSi$_2$O$_6$Cl$_2$ molecule. }
    \label{fig:dummy}
  \end{figure}
\end{document}